\documentclass[useAMS,usenatbib]{mn2e}
\usepackage{graphicx}
\usepackage{hyperref}
\usepackage{color}
\newcommand{\msun}{\mbox{${\rm M}_{\odot}$}}
 \newcommand{\btxt}[1]{{#1}}

 \newcommand{\boldtext}[1]{{#1}}

\title[Scaling relations from CFHTLenS]{CFHTLenS: Weak lensing calibrated
  scaling relations for low mass clusters of galaxies}

\author[Kettula et al.]{K. Kettula$^{1,2}$\thanks{E-mail: kimmo.kettula@helsinki.fi}, S. Giodini$^3$, E. van Uitert$^4$, H. Hoekstra$^3$, A. Finoguenov$^{1,5}$, \newauthor
M. Lerchster$^{1,5}$, T. Erben$^{4}$, C. Heymans$^{6}$, H. Hildebrandt$^{4,7}$, T. D. Kitching$^{8}$, \newauthor 
A. Mahdavi$^{9}$, Y. Mellier$^{10,11}$, L. Miller$^{12}$,  M. Mirkazemi$^{5}$, L. Van Waerbeke$^{7}$,  \newauthor 
J. Coupon$^{13,14}$,   E. Egami$^{15}$, L. Fu$^{16}$, M. J. Hudson$^{17,18}$, J. P. Kneib$^{19,20}$, K. Kuijken$^{3}$, \newauthor 
H. J. McCracken$^{10,11}$, M.~J. Pereira$^{14}$, B. Rowe$^{21,22}$, T. Schrabback$^{3,4,23}$, M. Tanaka$^{24}$, \newauthor
M. Velander$^{3,10}$ \\
$^1$~University of Helsinki, Department of Physics, P.O. Box 64, FI-00014 University of Helsinki, Finland \\
$^2$~Helsinki Institute of Physics, P.O. Box 64, FI-00014 University of Helsinki, Finland \\
$^3$~Leiden Observatory, Leiden University, PO Box 9513, 2300 RA, Leiden, the
Netherlands \\
$^4$~Argelander-Institut f\"ur Astronomie, Auf dem H\"ugel 71, 53121 Bonn, Germany \\
$^5$~Max-Planck-Institut f\"ur Extraterrestrische Physik, Giessenbachstraße, D-85740 Garching, Germany\\
$^{6}$~Scottish Universities Physics Alliance, Institute for Astronomy, University of Edinburgh, Royal Observatory, Blackford Hill, Edinburgh, EH9 3HJ, UK.\\
$^{7}$~Department of Physics and Astronomy, University of British Columbia, 6224 Agricultural Road, Vancouver, V6T 1Z1, BC, Canada.\\  
$^{8}$~Mullard Space Science Laboratory, University College London, Holmbury St Mary, Dorking, Surrey RH5 6NT, UK.\\
$^{9}$~Department of Physics and Astronomy, San Francisco State University, 1600 Holloway Avenue, San Francisco, CA 94132, USA \\
$^{10}$~Sorbonne Universités, UPMC Univ Paris 06, UMR 7095, Institut d’Astrophysique de Paris, F-75014, Paris, France \\
$^{11}$~CNRS, UMR 7095, Institut d’Astrophysique de Paris, F-75014, Paris, France \\
$^{12}$~Department of Physics, Oxford University, Keble Road, Oxford OX1 3RH, UK.\\ 
$^{13}$~Astronomical Observatory of the University of Geneva, ch. d’Ecogia 16,1290 Versoix, Switzerland \\
$^{14}$~Institute of Astronomy and Astrophysics, Academia Sinica, P.O. Box 23-141, Taipei 10617, Taiwan \\
$^{15}$~Steward Observatory, University of Arizona, 933 North Cherry Avenue, Tucson, AZ 85721, USA \\ 
$^{16}$~Shanghai Key Lab for Astrophysics, Shanghai Normal University, 100 Guilin Road, 200234, Shanghai, China \\
$^{17}$~Department of Physics and Astronomy, University of Waterloo, 200 University Avenue West, Waterloo, ON N2L 3G1, Canada \\
$^{18}$~Perimeter Institute for Theoretical Physics, 31 Caroline Street N, Waterloo, ON, N2L 1Y5, Canada.\\
$^{19}$~Laboratoire d'Astrophysique de Marseille, CNRS-Universit\'e, Pôle de l'Etoile Site de Ch\^ateau-Gombert 38, rue Fr\'ed\'eric Joliot-Curie, F-1338 \\
$^{20}$~Laboratoire d’Astrophysique EPFL, Observatoire de Sauverny, Versoix 1290, Switzerland \\
$^{21}$~Department of Physics and Astronomy, University College London, Gower Street, London WC1E 6BT, UK \\
$^{22}$~California Institute of Technology, 1200 E California Boulevard, Pasadena CA 91125, USA \\ 
$^{23}$~Kavli Institute for Particle Astrophysics and Cosmology, Stanford University, 382 Via Pueblo Mall, Stanford, CA 94305-4060, USA.\\
$^{24}$~National Astronomical Observatory of Japan 2-21-1 Osawa, Mitaka, Tokyo, 181-8588, Japan \\
}

\begin{document}
\date{Dates}

\newpage

\pagerange{\pageref{firstpage}--\pageref{lastpage}} \pubyear{2015}

\maketitle
\label{firstpage}

\clearpage

\begin{abstract}
\boldtext{We present weak lensing and X-ray analysis of 12 low mass clusters from the CFHTLenS and XMM-CFHTLS surveys. We combine these systems with high-mass systems from CCCP and low-mass systems from COSMOS to obtain a sample of 70 systems, spanning over two orders of magnitude in mass. We measure core-excised L$_X$-T$_X$, M-L$_X$ and M-T$_X$ scaling relations and include corrections for observational biases. By providing fully bias corrected relations, we give the current limitations for L$_X$ and T$_X$ as cluster mass proxies. We demonstrate that T$_X$ benefits from a significantly lower intrinsic scatter at fixed mass than L$_X$. By studying the residuals of the bias corrected relations, we show for the first time using weak lensing masses that galaxy groups seem more luminous and warmer for their mass than clusters. This implies a steepening of the M-L$_X$ and M-T$_X$ relations at low masses. We verify the inferred steepening using a different high mass sample from the literature and show that variance between samples is the dominant effect leading to discrepant scaling relations. We divide our sample into subsamples of merging and relaxed systems, and find that mergers may have enhanced scatter in lensing measurements, most likely due to stronger triaxiality and more substructure. For the L$_X$-T$_X$ relation, which is unaffected by lensing measurements, we find the opposite trend in scatter. We also explore the effects of X-ray cross-calibration and find that Chandra calibration leads to flatter L$_X$-T$_X$ and M-T$_X$ relations than XMM-Newton.}
\end{abstract}

\begin{keywords}
cosmology: observations -- cosmology: dark matter -- Physical data and process: gravitational lensing: weak --
galaxies: clusters: general
\end{keywords}

\section{Introduction}
Precise knowledge of the total mass of galaxy clusters is a crucial
ingredient in order to probe cosmology by means of cluster number
counts.  Cluster masses can be inferred by means of
gravitational lensing, from the velocity dispersion of cluster
galaxies assuming dynamical equilibrium, or from X-ray surface brightness and
temperatures assuming hydrostatic equilibrium (HSE).  However, these
direct methods are observationally expensive, especially for low-mass
systems and at high redshifts. Fortunately, cluster mass scales with
observational properties such as X-ray luminosity and temperature.
Therefore it is possible to calibrate robust and well understood
scaling relations between cluster mass and observables, in order to be
able to study statistical samples of clusters as cosmological probes.

\btxt{Both simulations and observations show that clusters are found in
various dynamical states, with bulk motions and non-thermal pressure
components present in the intracluster gas. These affect
mass measurements relying on dynamical equilibrium or HSE. In
particular, as indicated in both simulations 
\citep[e.g.][]{Nagai07,Shaw10,Rasia12}, observations 
\citep[e.g.][]{Mahdavi08,Mahdavi13,Kettula13,Donahue14,Israel14i,Israel14ii,vdl14} 
and recent analytical work by \citet{Shi14},
HSE mass estimates differ from the lensing mass. 
The trend in the above studies is that HSE mass estimates underestimate 
the true mass by $\sim$ 10--30 \%. However, as shown by e.g. the recent systematic comparison 
of mass estimates by \citet{Sereno14}, there is significant disagreement between different mass 
estimates relying on the same method.
Though cluster triaxiality and substructure may complicate the interpretation, 
gravitational lensing provides the most reliable way of
determining the true cluster mass, as it requires no assumptions on 
the thermodynamics of the
intracluster gas or the dynamical state of the cluster.
}

\btxt{In the self-similar case which assumes pure gravitational heating, cluster observables and mass are related by power-laws 
\citep{Kaiser86}.  However, the relative strength of baryonic physics increases at low masses.
Analysis by e.g. \citet{Nagai07,Giodini10,Mccarthy10,Stanek10,Fabjan11,LeBrun14,Planelles14,Pike14} indicate that 
baryonic processes such as non-gravitational feedback from star formation and active galactic nuclei 
(AGN) activity are expected to bias scaling relations from the self-similar prediction. The above works also indicate that the deviations are 
expected to be stronger for groups and low-mass clusters than for high-mass clusters.
Hydrodynamical simulations by \citet{Schaye10} show that the gas removed by AGN activity in groups can also 
affect the large scale structure out to several Mpc, potentially skewing cosmic shear measurements \boldtext{\citep{vanDaalen11,Semboloni11,Semboloni13,Kitching14}}. 
Consequently, characterisation of the effects of feedback at group and low-mass cluster level is of 
high interest for both cluster and cosmic shear studies.  
}

\btxt{Indeed, recent detailed observations of groups and low-mass clusters
by e.g. \citet{Sun09}, \citet{Eckmiller11} and \citet{Lovisari15} have reported evidence pointing to 
the direction of such mass dependent deviations from self-similar scaling \citep[see also][and references therein]{Giodini13}.
Even if a direct measurement of a break in the scaling relations is hard, relations fitted to groups tend have a larger intrinsic scatter 
than similar relations fitted to massive clusters. 
However, most previous studies rely on X-ray mass estimates based on HSE. The HSE condition is broken by the same feedback processes affecting 
the scaling relations, and HSE masses are thus likely strongly biased for
these low-mass systems \citep{Kettula13}. Therefore mass measurements
by means of gravitational lensing are instrumental at group and low
mass cluster scales.
}

\btxt{In the weak lensing regime the gravitational potential of the cluster distorts light
emitted by a background galaxy, resulting in a modified source
ellipticity, known as shear. As galaxies have an intrinsic ellipticity which is
typically larger than the lensing induced shear but not aligned with relation to 
the cluster, the shear has to be averaged over a statistical
sample of source galaxies in order to measure the weak lensing signal.}

\begin{figure}
\label{fig:balance}
\leavevmode \hbox{%
\includegraphics[width=\columnwidth]{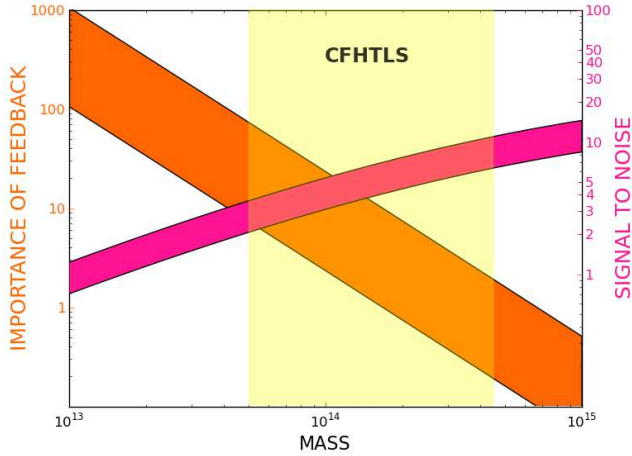}}
\caption{The importance of feedback (in orange) increases in
  systems of lower mass since the balance between the gravitational
  forces and the energetic processes happening in the core of galaxies
  (mostly linked to massive black holes) changes in favour of the
  latter \citep{Giodini10}. The signal-to-noise of weak lensing observations (in
  magenta) determining how well we can measure the total mass of the
  system, increases for systems of larger mass. These opposite behaviours
  define a "sweet spot" in the mass range at 10$^{14}$\msun, where
  feedback is important and the mass of individual
  systems is measurable with weak lensing. 
  With the CFHTLS we can study systems exactly in this mass
  range (yellow shaded area).}
\end{figure}

\btxt{The scaling of weak lensing mass to X-ray observables at galaxy group levels has previously
only been studied in the COSMOS field by \citet{Leauthaud10} and
\citet{Kettula13}, and recently at low-mass cluster levels by \citet{Connor14}.}
In this work we focus on studying the scaling of
weak lensing mass to X-ray luminosity L$_X$ and spectroscopic temperatures T$_X$
for a sample of low-mass clusters, with a typical mass of $\sim
10^{14}$ \msun. The studied systems are in the \btxt{"sweet spot"}, where they are massive enough 
to be studied with reasonable observational effort and, at the same time, non-gravitational
processes still give a significant contribution to their energetics (see Fig. \ref{fig:balance}). 
 This is quantified in Figure 1, which
shows the ratio of non-gravitational mechanical energy released by AGNs
to the gravitational binding energy of the intracluster gas and the weak 
lensing signal-to-noise ratio as a function of cluster mass. The ratio of 
the mechanical and binding energy is the average relationship from Fig. 1 in 
\citet{Giodini10}, the weak lensing signal-to-noise is based on \citet{Hamana04}. 

We use lensing measurements of individual systems from the Canada-France-Hawaii Telescope
Lensing Survey (CFHTLenS) and XMM-Newton X-ray observations from the
XMM-CFHTLS survey. \btxt{We refer to this sample as CFHTLS in this paper.
This sample also includes one system from the XMM-LSS survey. } 
\boldtext{We also include lower mass systems from COSMOS 
\citep{Kettula13} and massive clusters from CCCP \citep{Hoekstra12,Mahdavi13,Hoekstra15}
in order to study the mass dependence of the scaling relations. Combining the data from these three surveys allows us constrain 
weak lensing calibrated scaling relations using a long mass baseline spanning approximately two orders of magnitude. }

\boldtext{As pedagogically illustrated in Appendix A of \citet{Mantz10} scaling relations are affected by both Malmquist and Eddington bias. Malmquist bias will only affect the relations in case of covariance between the intrinsic scatters of the observable used for cluster detection and the measurables under investigation. However, the effect of Eddington bias can not be eliminated in the presence of intrinsic scatter about the mean relation \citep{Eddington13} - because of the interplay between the steep decline at high masses of the mass function and intrinsic scatter of luminosity and temperature, it is more likely that lower mass systems scatter towards a higher luminosity or temperature, than vice versa. This renders massive clusters hotter and more luminous for their mass than intermediate-mass systems, whereas this is less of an issue for the low and intermediate-mass samples, where the mass function is flatter. In order to understand the mass dependence of the scaling relations, the effect of observational biases have to be considered. As shown by e.g. \citet{Rykoff08} and \citet{Mantz14}, these effects can be modelled. }

\boldtext{Clusters typically undergo several mergers during their formation, 
leading to a varying degree of substructure and triaxial asymmetry. As our sample contains only 
measurements of individual systems, we are able to study the effects of the merger and residual 
activity on the scaling relations by dividing our sample into subsamples of relaxed  and 
non-relaxed systems by the amount of substructure.}

\boldtext{Finally, galaxy cluster measurements are affected by cross-calibration uncertainties of X-ray detectors. This has been shown by the International Astronomical Consortium for High Energy Calibration IACHEC \footnote{\url{http://web.mit.edu/iachec/}}  \citep{Nevalainen10,Kettula13b,Schellenberger14}, and independently by e.g. \citet{Snowden08}, \citet{Mahdavi13}, \citet{Donahue14} and \citet{Israel14ii}. These studies indicate that cluster temperatures measured with the Chandra observatory are typically $\sim$ 10--15 \% higher than those measured with XMM, whereas luminosities tend to agree to a few per cent. By investigating stacked residuals, the reported discrepancies can be accounted for by differences in the energy dependence of the effective area \citep{Kettula13b,Schellenberger14,Read14}. }

The lensing measurements are presented in Section
\ref{sec:cfhtlens} and X-ray observations in Section \ref{sec:xmm}. We
derive the lensing masses in Section~\ref{sec:shear} and present the scaling
relations between lensing mass and X-ray luminosity and temperature in
Section~\ref{sec:scaling}. \boldtext{We include bias corrections, and
study the effects of cluster morphology and X-ray cross-calibration.} 
Finally, we discuss our results in Section \ref{sec:disc}, 
and summarise our work and present our conclusions in Section \ref{sec:sum}. 
We denote scaling relations as Y-X, with Y as the dependent variable (y-direction) 
and X as the independent variable (x-direction).
We assume a flat $\Lambda$CDM cosmology
with H$_0$ = 72 km s$^{-1}$ Mpc$^{-1}$, $\Omega_M = $ 0.30 and
$\Omega_{\Lambda}$ = 0.70. All uncertainties are at 68\%
significance, unless stated otherwise.

\section{Data}

\subsection{The Canada-France-Hawaii Telescope Lensing Survey}
\label{sec:cfhtlens}

The CFHT Lensing Survey (CFHTLenS) is based on the Canada-France-Hawaii Telescope
Legacy Survey (CFHTLS), where a total area of 154 deg.$^2$ was imaged in 5
optical bands ($u^{*}g'r'i'z'$). The data are spread over four distinct
contiguous fields. The Northern field W3 ($\sim$ 44.2 deg.$^2$) lacks
X-ray coverage, but large fractions of the three equatorial fields
(W1: $\sim$64 deg.$^2$;W2: $\sim$23 deg.$^2$;W4: $\sim$23 deg.$^2$)
were observed by {\it XMM-Newton} as part of the XMM-CFHTLS survey
(Section \ref{sec:xmm}).

The deep, multi-colour data enable the determination of photometric
redshifts of the sources \citep{Hildebrandt12} which are used to
improve the precision of the lensing mass estimates by taking
advantage of the redshift dependence. The $i'$-band data, which reach
$i_{AB}=25.5$ (5$\sigma$), are used for the lensing measurements
because of the excellent image quality. To determine an accurate
lensing signal from these data also requires a special purpose
reduction and analysis pipeline which was developed and tested by
us and is described in detail in \cite{Heymans12,Erben13}. We
discuss some of the key steps in the weak lensing analysis, but refer
the interested reader to the aforementioned CFHTLenS papers for a more
detailed discussion.

A critical step in the weak lensing analysis is the accurate
measurement of galaxy shapes. As the CFHT data consist of multiple $i'$-band exposures (typically seven), the
algorithm needs to be able to account for the varying PSF between
exposures.  The Bayesian fitting code {\tt lensfit} \citep{Miller07,Miller13}
was used for this purpose. The resulting
catalog\footnote{http://cfhtlens.org/astronomers/data-store} includes
measurements of galaxy ellipticities, $\epsilon_{1}$ and
$\epsilon_{2}$, which can be used as estimators of the shear with an
inverse variance weight $w$.  Image simulations were used to determine
additional empirical shear calibration corrections, which depend on
signal-to-noise and galaxy size. These are described in
\cite{Miller13} and \cite{Heymans12}. These papers also present a
number of tests to identify residual systematics. A key test is the
measurement of the correlation between the PSF orientation and the
corrected galaxy shape.  \cite{Heymans12} found that 75$\%$ of the
data pass this test and thus can be used in the cosmological analyses
\citep{Kilbinger13, Benjamin13, Heymans13, Simpson13,Kitching14}.

Cosmic shear studies are very sensitive to such residual
correlations. In this paper, however, we measure the ensemble
azimuthally averaged signal around a large number of low mass
clusters. As is the case for the study of the lensing signal around
galaxies \citep{Hudson13,Velander13}, this measurement is much more robust
against residual (additive) biases. Therefore we follow
\cite{Velander13} and use all CFHTLenS fields in our analysis. Six of our clusters 
reside within 5 arcmin of the image edges. As the PSF varies across the field-of-view, 
it is different from the central and outer regions of a pointing. As an additional sanity check 
of the reliability of our cluster masses, we therefore compare the masses of these six clusters to the 
other ones. We do not find any systematic difference with respect to the scaling relations. 

\cite{Hildebrandt12} present measurements of the photometric redshifts
for the sources using the Bayesian photometric redshift code {\tt BPZ}
\citep{Benitez00}. Importantly, the PSF was homogenized between the
five optical bands, which improves the accuracy of the photometric
redshifts across the survey. The robustness of the photometric redshifts
was tested in \cite{Hildebrandt12} and \cite{Benjamin13}.

To ensure that robust shape measurements and reliable redshift
estimates are available, we limit the source sample to those with
$0.2<z_{\rm BPZ}<1.3$ and $i'<24.7$. The selection yields a scatter in
photometric redshift in the range $0.03<\sigma<0.06$ with outlier
rates smaller than $10\%$ \citep{Hildebrandt12}. We also exclude
galaxies that have the flag MASK $>$ 0 as their photometry and shape
measurement may be affected by image artifacts.  The resulting sample
has a weighted mean source redshift of $\langle z\rangle=0.75$ and an
effective number density of $n_{\rm eff}=11$ arcmin$^{-2}$.

\subsection{The XMM-CFHTLS survey}
\label{sec:xmm}

Eleven clusters with X-ray flux significance greater than 20,
corresponding to a minimum of 400 photons sufficient for reliable 
temperature measurements, have been observed by
XMM-Newton as a part of the XMM-CFHTLS survey (PI: Finoguenov, see \citet{Mirkazemi15}). 
\boldtext{We also include one cluster (XID102760) from the CFHTLS W1 field which has been observed as a part of 
the XMM-LSS survey, with the analysis presented in \citet{Gozaliasl14}.}
The clusters have been identified from ROSAT All Sky Survey data, through optical filtering 
using CFHTLS multiband data and spectroscopic follow-up with HECTOSPEC/MMT \boldtext{\citet{Mirkazemi15}.}

When compared to existing samples of galaxy clusters and groups,
XMM-CFHTLS covers an interesting range of properties, bridging the
intermediate mass range between groups and clusters. Because of the
combination of a wide area with a moderately deep X-ray coverage,
XMM-CFHTLS contains more low mass systems at intermediate redshift
than other XMM cluster samples such as REXCESS \citep{Bohringer07} or
LocuSS \citep{Smith05}, but not as low mass as those in COSMOS
\citep{Scoville07}. The typical system in XMM-CFHTLS is a low mass
cluster with a mean total mass of $\sim$10$^{14}$ \msun, so that we
can call these Virgo-sized systems.

\boldtext{In order to efficiently find the clusters in the full area of the CFHTLS survey, 
we used RASS sources and identify them using CHFTLS photometric data and studied 
their masses using the combination of shape measurements and photometry.
This X-ray selection  of
clusters for the scaling relation studies introduces a bias to the resulting scaling relation.}
The straightforward application is in using exactly the same quantity
that has been used in the selection, which is a \boldtext{total X-ray luminosity $L$}. Although
we do not include the scaling relation with total \boldtext{$L$} in this study,
it is important to mention that the calculation of bias needs to be
modified to account for the Eddington bias associated with the detection of
sources in RASS data. 
\btxt{The flux limit of the RASS data is formally 10$^{-13}$ erg s$^{-1}$ cm$^{-2}$ in a 0.5-2 keV energy band, 
corresponding to 4 counts. A number of systems with a mean expected number of counts below the RASS limit of 4 that 
have been upscattered to over 4 are expected to be selected as well.
For the scaling relations this leads to a {\it reduction} of bias.
Following the formulation of \citet{Vikhlinin09} we
can write the bias correction as
\begin{equation}
\label{eq:bias}
b(\ln L_o) = \frac{\int_{-\infty}^{+\infty} (\ln L - \ln L_0) P(T|C(\ln L,z)) e^{\frac{(\ln L - \ln L_o)^2}{2 \sigma^2}} d \ln L}{\int_{-\infty}^{+\infty} P(T|C(\ln L,z)) e^{\frac{(\ln L - \ln L_o)^2}{2 \sigma^2}} d \ln L }
\end{equation}
where $T$ is the RASS count threshold, $C(x,z)$ are the predicted RASS
counts from a cluster at a redshift $z$ with luminosity \boldtext{$L$},
$P(T|C(L,z))$ is the probability of detection, $\sigma$ is the scatter
of the scaling relation. The  bias for the average flux of the sources at the detection limit is 1.5 counts , 
leading to an average limit of $7 \times 10^{-14}$ erg s$^{-1}$ cm$^{-2}$, which is lower than the nominal 
RASS flux limit. XMM-Newton follow up removes this uncertainty from the flux and confirms the effect. 
For  bias calculation due to the flux limit for a putative survey with high statistics, the Poisson term should be 
replaced by a Gaussian around the flux limit. Most known clusters (e.g. REFLEX, NORAS, MACS), however, are selected 
from RASS down to count limits where Poisson effects are important. In this case Eq.\ref{eq:bias} should be used.  }

\btxt{The selection effects on the scaling relations involving other 
parameters than total luminosity depend on the covariance with the scatter. 
Since we work with \boldtext{core-excised temperature T$_X$ and luminosity L$_X$}, both
measured inside $0.1-1$ R$_{500}$ \footnote{the spherical overdensity 
radius inside which the density is 
500 times the critical density}, the bias due to selection on full
luminosity L can only be present if there is a covariance in the scatter
between the full luminosity and \boldtext{core-excised} T$_X$ and L$_X$.
For example if cool core clusters have slightly different properties in the outskirts, 
some residual bias might be present \citep{Zhang11}. 
However, at present the evidence for this effect is very marginal and we have decided not to correct for it.
By determining the scaling relations separately for relaxed and unrelaxed clusters, 
we remove the effects of such residual biases.}

\btxt{For calculating L$_X$ we used the full
aperture ($0.1-1$ R$_{500}$) and the measured temperature for K-correction, 
reducing the scatter associated with the assumption of the
shape of the emission and predicting temperatures using the L$_X$-T$_X$ relation. 
As X-ray selection preferentially detects relaxed clusters
(due to cool cores) and the gas distribution generally displays stronger spherical 
symmetry than the underlying dark matter distribution, we did not consider orientation 
dependence in cluster selection. As we expect the contribution from triaxiality to be minimal, 
we assume spherical symmetry. We study the validity of this assumption is Section \ref{sec:subst}.}

\btxt{In measuring the temperature we only use data from the EPIC-pn instrument, and performed a
local adjustment of the background in addition to the use of stored
instrument background,} as in \citet{finoguenov05, pratt07}, since the
clusters occupy only a small part of the detector.  In the spectral
analysis, we used the 0.5--7.5 keV energy band, excluding the 1.4--1.6 keV
interval affected by instrumental line emission. \boldtext{We used SAS version 13.5.0 
and corresponding calibration files to construct the responses.}

\section{Weak lensing signal}
\label{sec:shear}

The differential deflection of light rays by an intervening lens leads
to a shearing (and magnification) of the images of the
sources \citep[see e.g.][for a recent review on gravitational lensing
studies of clusters]{Hoekstra13}. The resulting change in ellipticity,
however, is typically much smaller than the intrinsic source
ellipticity and an estimate for the shear is obtained by averaging the
shapes of an ensemble of source galaxies.

As the survey volume increases, the massive systems are found at
higher redshift. Unfortunately, the lensing signal decreases as the
lens approaches the source redshift. This is because the amplitude of
the lensing signal is inversely proportional to the critical surface
density $\Sigma_{\rm crit}$ given by

\begin{equation}
\Sigma_{\rm crit}=\frac{c^2}{4\pi G}\frac{D_s}{D_l D_{ls}},
\end{equation}

\noindent where $D_l$ is the angular diameter distance to the lens,
$D_s$ the angular diameter distance to the source, and $D_{ls}$ the
angular diameter distance between the lens and the
source. 

Hence the redshift dependence of the lensing signal and the noise due
to the intrinsic shapes of the finite number of sources, limit both
the mass and redshift range for which individual cluster masses can be
measured.  To ensure a sufficient number density of background galaxies we limit
the analysis to clusters with $z<0.6$.

\begin{figure}
\includegraphics[width=\columnwidth]{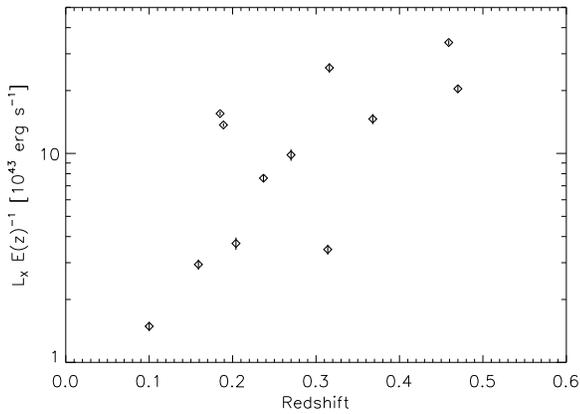}
\caption{X-ray luminosity versus redshift for our cluster sample selected from XMM-CFHTLS 
\citep{Mirkazemi15}.
}
\label{fig:lx_z}
\end{figure}

To determine the mass, it is convenient to azimuthally average
the tangential shear $\langle\gamma_T\rangle$ as a function of radius
from the lens, and fit a parameterized model to the signal. The {\tt
  lensfit} measurements yield ellipticities $\epsilon_1$ and
$\epsilon_2$, and the tangential shear is the projection perpendicular
to the direction (with azimuthal angle $\phi$) connecting the source
galaxy and the lens. It is given by

\begin{equation}
\gamma_T= -(\epsilon_{1}~ \cos(2\phi)+\epsilon_{2}\times \sin(2 \phi)).
\end{equation}

\noindent It is also convenient to measure the cross-shear

\begin{equation}
\gamma_X= -(\epsilon_{1}~ \sin(2\phi)-\epsilon_{2}\times \cos(2 \phi)),
\end{equation}

\noindent whose azimuthal average is expected to vanish in the absence of systematic
effects and is therefore used as a diagnostic. Note that we assume
that the images are oriented randomly in the absence of lensing.
Although this assumption may not hold in general \citep[see e.g.][]{Heymans13}, the amplitude
is found to be small, but also it should not contribute to
the tangential shear around lenses.

As discussed in Section \ref{sec:cfhtlens} we only use sources with $i'<24.7$,
to ensure a robust shape measurement and we limit our sample to
$0.2<z<1.3$, to ensure the robustness of the photometric redshifts
\citep{Hildebrandt12}. To minimize the contamination of cluster members in our source sample, we consider only source galaxies with a photometric redshift larger than $z_{\rm lens}$+0.15. The redshift cut of 0.15 is a conservative one, and results in negligible contamination of cluster galaxies in the source sample. Including sources even closer to the lens redshift would not lead to a large improvement in signal-to-noise, as their lensing efficiencies are small. As the redshifts of our clusters are $< 0.6$, the photo-z errors of the sources are almost flat close to the lens redshift \citep{Hildebrandt12}, and the photo-z cut needs not be redshift dependent.

Thus we sort the source galaxies in 15 equally sized radial bins from 0.15 Mpc from the center
of the lens (in our case the low-mass cluster) out to a radius of
3~Mpc. We define the center as the location of the X-ray peak. 
In each bin we perform a weighted average of the lensing signal as:
\begin{equation}
\langle\Delta\Sigma\rangle(r)=\frac{\sum{w_{i}\Sigma_{{\rm crit},i}\gamma_{T,i}(r)}}{\sum w_{i}},
\label{aveprof}
\end{equation}
where the lensing weight $w_i$ quantifies the quality of the shape measurement \citep[see][for details]{Miller13}. We compute $\Sigma_{{\rm crit},i}$ by integrating over the redshift distribution of each source galaxy.
Secondly, we apply a weight of $\Sigma_{\rm crit}^{-2}$ to each lens-source pair, effectively down-weighing source galaxies that are close in redshift to the lens. As mentioned
in Section \ref{sec:cfhtlens} the {\tt lensfit} output ellipticities need to
be corrected for a multiplicative bias that depends on signal-to-noise
and size $m(\nu_{\rm SN},r_{\rm gal})$. As discussed in
\cite{Miller13}, simply dividing the shear for each galaxy by a factor
$(1+m)$ would lead to a biased estimate of the average. Instead we
compute the corrected shear as follows:

\begin{equation}
\langle\Delta\Sigma^{\rm cor}\rangle(r)=\frac{\langle\Delta\Sigma\rangle(r)}{1+K(r)},
\end{equation}

\noindent where the correction is given by

\begin{equation}
1+K(r)=\frac{\sum{w_i[1+m(\nu_{\rm SN},r_{\rm gal})]}}{\sum w_i},
\end{equation}
with $\nu_{\rm SN}$ \btxt{stands for} the signal-to-noise ratio of the galaxy and $r_{\rm gal}$ the size. The error on the shear signal is computed by taking the inverse square root of the sum of the weights, and accounts for intrinsic shape noise as well as measurement noise.

To estimate cluster masses, we assume that the matter
density is described by an NFW profile \citep{NFW97}, which is found
to be a good approximation to simulated profiles in N-body simulations
of collisionless cold dark matter.  The density profile is given by

\begin{equation} \label{eq:nfwprofile} 
  \rho(r) = \frac{\delta_{c}
  \rho_{\rm crit}}{\frac{r}{r_{s}} \left( 1 + \frac{r}{r_{s}} \right)^{2}},
\end{equation}

\noindent where $\rho_{\rm crit}=3 H^2(z)/8 \pi G$ is the critical
density of the universe at the lens redshift $z$ and $H(z)$ is the
corresponding Hubble parameter.  The scale radius $r_s$ is related to
the virial radius $r_{\rm vir}$ by the concentration parameter $c_{\rm
  vir}=r_{\rm vir}/r_s$ and $\delta_{c}$ is related to $c_{\rm
  vir}$ by 
\begin{equation}
\delta_{c} = \frac{\Delta_{\rm vir}}{3} ~\frac{c_{\rm vir}^3}{\ln (1 + c_{\rm vir}) - \frac{c_{
\rm vir}}{1 + c_{\rm vir}}},
\end{equation}
where $\Delta_{\rm vir}$ is the average overdensity inside $r_{\rm vir}$. 
Alternatively we can express the mass in
terms of $M_\Delta$, the mass contained within a radius $r_\Delta$
where the mean mass density is $\Delta \times \rho_{\rm
  crit}$. Results are commonly listed for $\Delta=200$ and $\Delta=500$.

\begin{figure*}
\begin{center}
\includegraphics[angle=270,width=\textwidth]{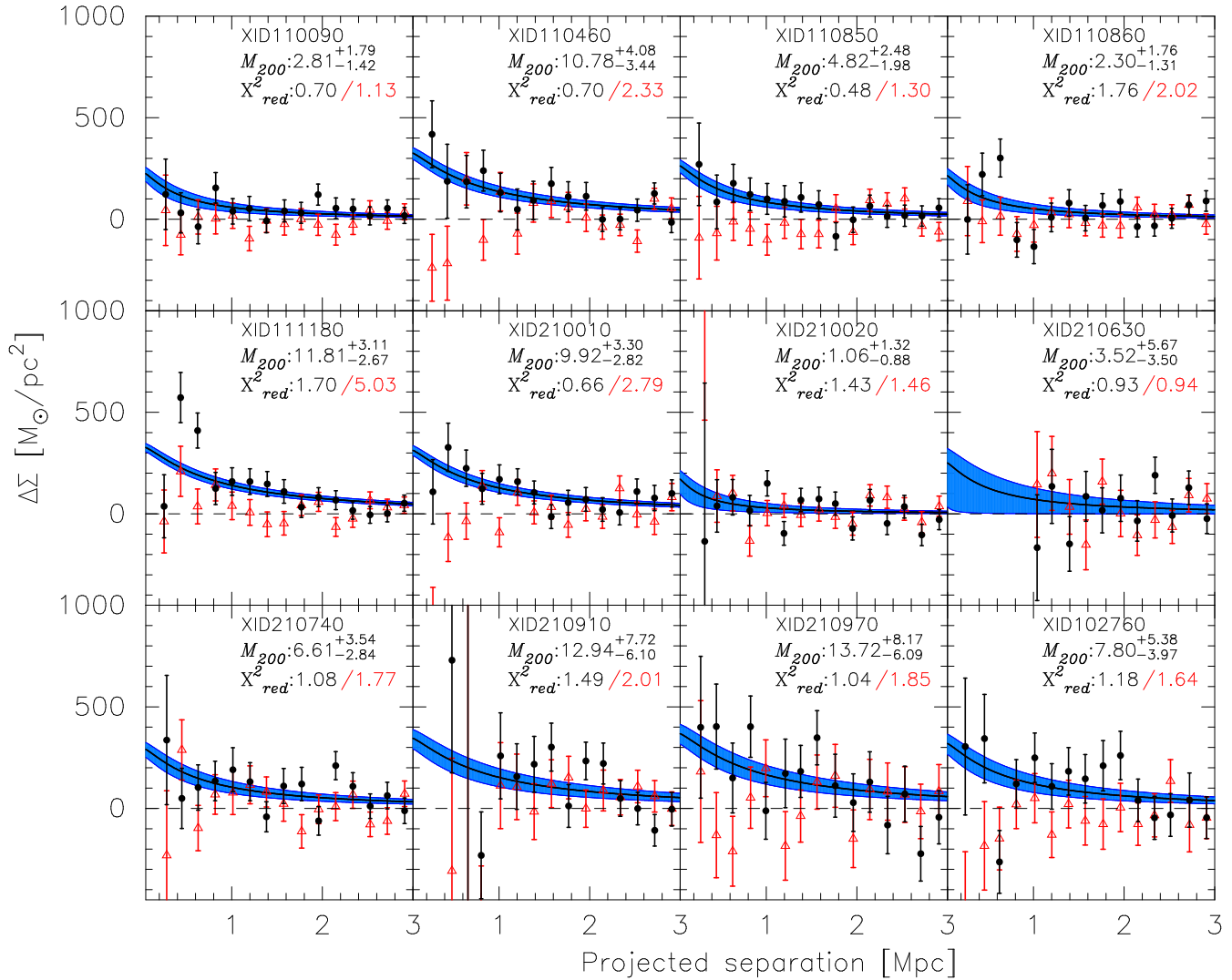}
\caption{Shear profiles out to 3~Mpc for the individual X-ray clusters measured
  using CFHTLenS data that were detected with an X-ray flux significance higher than 20,
  corresponding to a minimum of 400 photons. The blue shaded line
  shows the uncertainty on the best fitting profile. Each panel shows the 
  mass M$_{200}$ and the error of the mass in units of $10^{14}$ \msun, measured shear 
  profiles and the $\chi^2$ values for the NFW profile fit to the tangential shear (black circles). 
  The cross-shear and the $\chi^2$ value of the null-hypothesis that the tangential shear signal is zero are shown in red.
  Cluster XID210640 falls in the middle of a large stellar halo mask and lacks data on smaller scales.
}
\label{fig:single}
\end{center}
\end{figure*}

Numerical simulations also indicate that the virial mass $M_{\rm vir}$
and the concentration are correlated, with more massive systems having
lower values for $c_{\rm vir}$. Here we use the results from
\cite{Duffy08}, which give

\begin{equation}
\label{eq:mc}
  c=5.71\times \left(\frac{M_{200}}{2\times10^{12}h^{-1}}\right)^{-0.084}
  \times(1+z)^{-0.47}.
\end{equation}

Analytic expressions for the tangential shear of NFW profiles have been derived by
\citet{Wright00} and \citet{1996A&A...313..697B}. We fit the NFW
model shear to the profiles shown in Fig. \ref{fig:single}
and indicate the best fit model by the solid line. The coloured region
indicates the $68\%$ region for the model. As we measure M$_{200}$ from the NFW profile 
using the mass--concentration relation in Eq. \ref{eq:mc}, we have one free parameter for 15 radial bins giving 14 degrees of freedom (we note that cluster XID210640 falls in the middle of a large stellar halo mask and lacks data on smaller scales).  
We test the best-fit NFW profile against the null hypothesis that the tangential shear signal is zero and show 
the reduced $\chi^2$ values in Fig. \ref{fig:single}.
We use the best-fit NFW profile to rescale virial mass to M$_{500}$.  The resulting values for
$M_{200}$ and $M_{500}$ are listed in Table~\ref{table1}.

These are indeed the most massive clusters in the XMM-CFHTLS data, but
the observed lensing signal is nevertheless quite sensitive to
contributions from uncorrelated large-scale structure along the
line-of-sight \citep{Hoekstra01b, Hoekstra11b} or substructure and triaxial 
shape of the cluster halo \citep{Corless07,Meneghetti10,Becker11}. Such structures modify the observed
tangential shear profile. Both effects are an additional source of noise, whereas 
the latter might lead to biased mass estimate if we fit an NFW model to the data. 

The $\chi^2$ values of the NFW profile fits shown in Fig. \ref{fig:single} show that the 
data are well described by a single NFW profile. However, we note that for XID210910 a secondary 
group is detected in the X-ray image, which would tend to bias the NFW mass high. 

\begin{table*}
\label{table1}
\caption{Table of X-ray measurements and weak lensing masses for systems in our sample}
\centering
\begin{tabular}{lcccccccc}
\hline
XID & RA & DEC & $z$ & L$_X$ & T$_X$ & $M_\mathrm{200}$ &$M_\mathrm{500}$  & D$_{\rm BCG}$ \\
    & deg & deg &   &10$^{43}$ erg sec$^{-1}$ & keV & 10$^{14}$ M$_{\odot}$ & 10$^{14}$ M$_{\odot}$ &  kpc \\
\hline
  110090 & 36.2713 & -9.8381 & 0.159 & 3.16$\pm$0.18  & 3.62$\pm$0.79 & $2.81^{+1.79}_{-1.42}$ & $2.00^{+1.28}_{-1.02}$ &   17 \\  
  110460 & 35.998  & -8.5956 & 0.27  & 11.19$\pm$0.71 & 7.25$\pm$3.19 & $10.78^{+4.08}_{-3.44}$ & $7.45^{+2.82}_{-2.38}$ &  28  \\ 
  110850 & 33.6064 & -6.4605 & 0.237 & 8.52$\pm$0.35  & 2.39$\pm$0.7  & $4.82^{+2.48}_{-1.98}$ & $3.38^{+1.74}_{-1.39}$ &   17  \\ 
  110860 & 36.3021 & -6.3837 & 0.204 & 4.0$\pm$0.28   & 3.87$\pm$1.19 & $2.30^{+1.76}_{-1.31}$ & $1.64^{+1.26}_{-0.93}$ &   13   \\ 
  111180 & 37.9269 & -4.8814 & 0.185 & 16.90$\pm$0.37 &  5.0$\pm$0.61 & $11.81^{+3.11}_{-2.67}$ & $8.23^{+2.17}_{-1.86}$ &   62  \\ 
  210010 & 133.0656 & -5.5651 & 0.189 & 14.94$\pm$0.29 & 4.88$\pm$0.62 & $9.92^{+3.30}_{-2.82}$ & $6.93^{+2.31}_{-1.97}$ &   24   \\ 
  210020 & 134.6609 & -5.4211 & 0.1   & 1.56$\pm$0.08  & 1.65$\pm$0.3  & $1.06^{+1.32}_{-0.88}$ & $0.77^{+0.96}_{-0.64}$ &   431  \\
  210630 & 133.5554 & -2.3499 & 0.368 & 17.53$\pm$0.98 & 5.31$\pm$2.48 & $3.52^{+5.67}_{-3.50}$ & $2.45^{+3.95}_{-2.44}$ &   29  \\ 
  210740 & 135.4147 & -1.9799 & 0.314 & 4.04$\pm$0.22  & 4.59$\pm$1.57 & $6.61^{+3.54}_{-2.84}$ & $4.58^{+2.45}_{-1.97}$ &   21 \\ 
  210910 & 135.3770 & -1.6532 & 0.316 & 29.95$\pm$1.56 & 5.04$\pm$2.42 & $12.94^{+7.72}_{-6.10}$ & $8.87^{+5.29}_{-4.18}$ &   30  \\ 
  210970 & 133.0675 & -1.0260 & 0.459 & 42.81$\pm$1.07  & 5.35$\pm$1.18 & $13.72^{+8.17}_{-6.09}$ & $9.25^{+5.50}_{-4.10}$ &   42 \\ 
  102760 & 35.4391  & -3.7712 & 0.47  & 25.88$\pm$1.13 &  8.2$\pm$5.55 & $7.80^{+5.38}_{-3.97}$ & $5.30^{+3.66}_{-2.70}$ &   32    \\ 
\hline
\end{tabular}

\medskip
XID is the X-ray identification number in the XMM-CFHTLS survey, RA and DEC are the coordinates of the cluster center defined by the X-ray peak, $z$ the redshift of the cluster, T$_X$ and L$_X$ the X-ray temperature and luminosity, $M_\mathrm{200}$ and $M_\mathrm{500}$ the spherical overdensity masses with respect to the critical density and D$_{\rm BCG}$ the offset between the brightest cluster galaxy and X-ray peak. 
\end{table*}

\subsection{Systematics in mass estimates}

\btxt{The accuracy of the scaling relations depends on the ability to measure unbiased cluster masses. In this section we investigate different 
systematic effects that can bias our lensing masses. 
}

\begin{figure}
\includegraphics[width=\columnwidth]{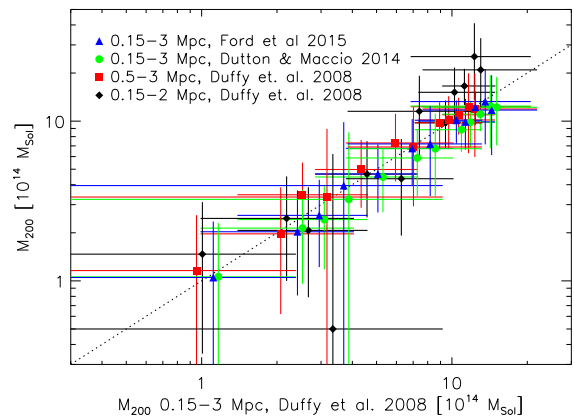}
\caption{\boldtext{Comparison of mass measurements assuming different mass-concentration relations, radial fit ranges or background galaxy filtering to the mass measurements adopted in this work. 
}}
\label{fig:sys}
\end{figure}

\btxt{As we fit the density profiles down to a radial range of 150 kpc, the resulting masses can be affected by  the mass--concentration relation assumed for the NFW profile. This was explored by \citet{Hoekstra12}, who showed that the sensitivity to the mass-concentration depends on the fit range and overdensity $\Delta$. 
They found their masses using a fit range of 0.5--2.0 Mpc to be most stable with $\Delta = 1000$. 
To investigate how sensitive our masses are to the selected mass--concentration relation we fit the NFW profiles 
assuming the relation of \citet{DM14}. We find that the average ratio of best-fit masses using \citet{DM14} to \citet{Duffy08} is 0.92
 $\pm$ 0.04, i.e. \citet{DM14} results on average in lower masses by 2$\sigma$ (see Fig. \ref{fig:sys}). 
 As an additional test, we also  measured our masses by excluding the central 0.5 Mpc 
 and find perfect agreement with our reported 
mass estimates. The average ratio of best-fit masses is 0.99 $\pm$ 0.11 (see Fig. \ref{fig:sys}). 
}

\btxt{Simulations by \citet{Becker11} suggest that extending the fit range beyond the virial radius may bias lensing masses low by 5--10 \% due to the correlated large scale structure. 
To test this we adopt an upper fit range of 2 Mpc. In this case we find that the average ratio of the best-fit masses is 1.15 $\pm$ 0.49.
If fitting beyond the virial radius would bias our mass estimates low, the ratio of the best-fit masses should be larger for low-mass systems with 
smaller virial radii than for massive clusters. We are not able to detect this trend in the data (see Fig. \ref{fig:sys}).   
}

\btxt{In the lensing measurement, we compute the mean lensing efficiency $\langle D_{\rm ls}/D_{\rm s} \rangle$ for each source by integrating over the full stacked photo-z posterior probability distribution P(z). Since the relation between lensing efficiency and redshift is non-linear, this could introduce a bias if the stacked P(z) is not a fair representation of the actual redshift distribution of the sources. To estimate its size, we consider a single lens-source pair. For the lens, we adopt a redshift of 0.2. For the source, we assume a redshift probability distribution that is representative for objects in CFHTLenS \citep[see][]{Hildebrandt12}, i.e. we describe the stacked P(z) by a gaussian with a mean of 0.7 and a standard deviation of 0.05, plus a second gaussian with a standard deviation of 0.5 (but with the same mean) that contains 7 \% of the total probability, to account for an outlier fraction of 7 \%. We compare the input $D_{\rm ls}/D_{\rm s}$ to the one that is averaged over the stacked P(z), and find that the latter is biased low by 1\%. Repeating the test for a lens at a redshift of 0.5 and a mean source redshift of 0.9, we find a similar bias. 

\boldtext{If not properly accounted for, dilution by foreground galaxies can bias the mass measurements. Using the P(z) modelling above, we compute a mass dilution by foreground galaxies of 3.5 \%. As a final test, we re-measure the masses using the same selection criteria for background galaxies as \citet{Ford15}, i.e. that the peak of the galaxy's P(z) is higher than the redshift of the cluster and that at least 90 \% of the galaxy's P(z) is at a higher redshift than the cluster. In this case we find that the best-fit masses are consistent with our measurements, with an average ratio of 0.97 $\pm$ 0.08 (see Fig. \ref{fig:sys}). We also note that in case our mass measurements would be significantly diluted by foreground galaxies, the expected ratio would be higher than unity.} 
}

\section{Scaling Relations}
\label{sec:scaling}
\begin{figure*}
\includegraphics[width=\textwidth]{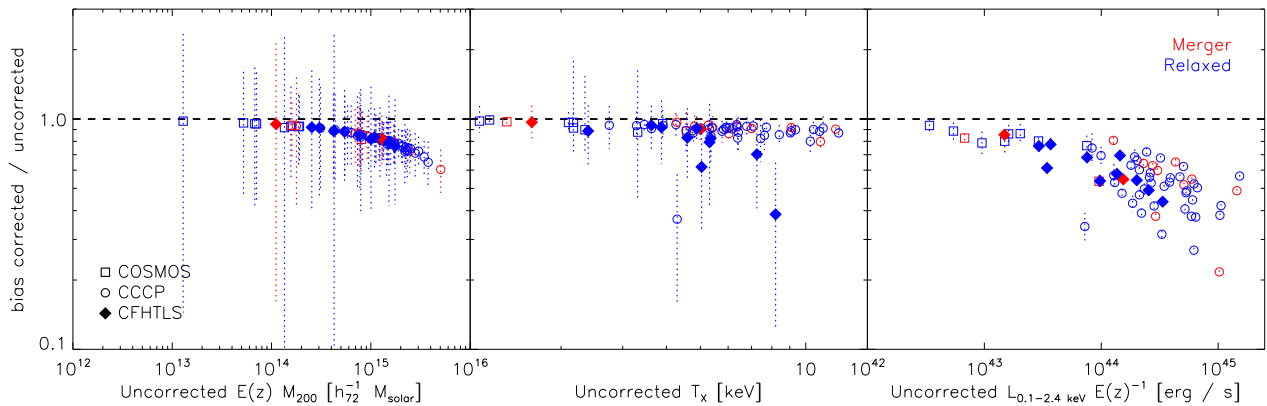}
\caption{\boldtext{The values of the Eddington bias corrections applied to mass (left panel), temperature (middle panel) and luminosity (right panel). Blue and red dotted data shows the residuals for individual merging and relaxed systems, squares indicate systems from COSMOS, circles from CCCP and solid diamonds from CFHTLS. Errors are the statistical errors of the measurements.}
}
\label{fig:bias}
\end{figure*}

The combination of X-ray and CFHTLenS weak lensing data is ideal for calibrating cluster mass proxies in the low-mass cluster regime. \boldtext{We present our fitting method, sample, bias corrections, and morphological classification of systems in Section \ref{sec:fit}. In Sections \ref{sec:lt}, \ref{sec:ml} and \ref{sec:mt} we present the scaling between weak lensing mass, core-excised X-ray luminosity and temperature,and discuss the global scaling properties (we explore the mass and morphology dependence of the relations in Sections \ref{sec:steep} and \ref{sec:subst}). Finally, we study the effects of X-ray cross-calibration in \ref{sect:calib}.}

\begin{figure}
\includegraphics[width=\columnwidth]{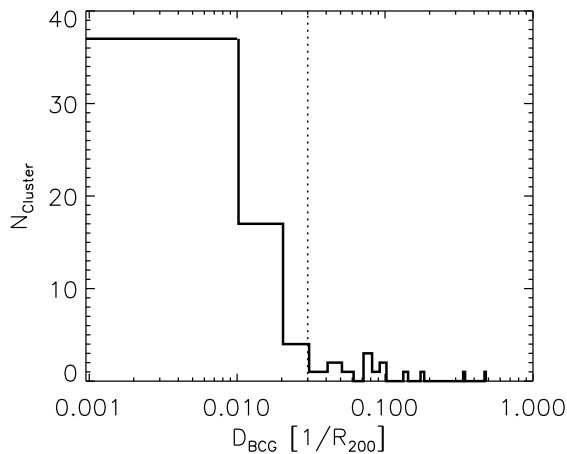}
\caption{The distribution of offsets between X-ray peak and BCG D$_{\rm BGC}$. D$_{\rm BGC}$ are given as fractions of R$_{200}$. The dotted vertical line separates between relaxed and merging clusters.
}
\label{fig:dbcg}
\end{figure}

\subsection{Fitting method}
\label{sec:fit}

The self-similar prediction for the scaling relation between two quantities A and B, such as mass and luminosity or luminosity and temperature, is a power-law , where the predicted value of slope $\alpha$ varies for the different relations \citep{Kaiser86}. Here we assume such a power-law form given by
\begin{eqnarray}
\label{eq:pl}
\log_{10} \frac{{\rm A \times E}(z)^{\rm n_A}}{\rm A_0} = \log_{10}(N) + \alpha \times \log_{10} \frac{{\rm B \times E}(z)^{n_{B}}}{\rm B_{0}} 
\end{eqnarray}

with A$_0$ and B$_0$ defining the pivot-point. E($z$) gives the scaling of overdensity with redshift and it is defined as 
\begin{equation}
\label{eq:ez}
{\rm E}(z) = \frac{H(z)}{H_0} = \sqrt{\Omega_{M}\,(1 + z)^3 + \Omega_{\Lambda}}.
\end{equation} 
$n_{\rm A}$ and $n_{\rm B}$ give the E($z$) dependence of quantities A and B. For mass $n_{\rm A}$ or $n_{\rm B}$ is 1, for L$_X$ it is -1 and for T$_X$ 0.

We let both the slope $\alpha$, normalization $\log_{10}(N)$ and intrinsic scatter  $\sigma_{\log({\rm A|B})}$ vary freely in the fits. \boldtext{We use the Bayesian linear regression routine of \citet{Kelly07} with the Metropolis-Hastings 
sampler to find the best-fit parameters. The routine includes intrinsic scatter \boldtext{in the dependent variable (i.e. y-direction)} $\sigma_{\log({\rm A|B})}$, which we expect to follow a log-normal distribution.} We define best-fit parameters as the median of the single parameter posterior distributions and errors as the values corresponding to the 68th percentiles. 

In order to improve the precision \boldtext{and to study the mass dependence} of the relation  we include measurements of 10 individual low-mass systems from the Cosmic Evolution Survey (COSMOS) and \btxt{48} individual high-mass systems from the Canadian Cluster Comparison Project (CCCP). \boldtext{We utilise the three surveys making up our sample as overlapping mass bins, with COSMOS forming the low-mass, CFHTLS intermediate-mass and CCCP the high-mass bin, and fit the scaling relations independently for each of the surveys.} 

COSMOS data, lensing and temperature measurements are presented in \citet{Kettula13}. The COSMOS systems have lensing masses based on deep HST imaging and 30+ band photometric redshifts, and X-ray measurements obtained with XMM-Newton. We derive luminosities from the COSMOS data using the method presented in Section \ref{sec:xmm} in this work \boldtext{(see Table \ref{tab:cosmos})}. 
\btxt{For the CCCP sample we use recent lensing mass measurements presented in \citet{Hoekstra15} measured assuming \boldtext{an NFW density profile and }the \citet{Duffy08} mass-concentration relation} and X-ray measurements obtained with both Chandra and XMM-Newton. \boldtext{We derive core-excised L$_X$ using the 0.1--2.4 keV band for the CCCP systems using the method described in \citet{Mahdavi13} \citep[see also][]{Mahdavi14} and use the core-excised temperatures from \citet{Mahdavi13}\footnote{Available on http://sfstar.sfsu.edu/cccp/}. The soft band L$_X$ measurements are given in Table \ref{tab:cccp}.}  Chandra observations of CCCP clusters are adjusted to match XMM-Newton calibration.  This gives us a sample of 72 individual systems, with T$_X$ $\sim$ 1--12 keV, L$_X$ $\sim 10^{43}$--10$^{45}$ erg/s and a mass from $\sim 10^{13}$ to a few times $10^{15}$ \msun.

We note that there are differences in the calibration of the lensing signal for these additional data sets, compared to \boldtext{CFHTLS}. Furthermore, the CCCP data lack photometric redshift information which may impact the correction for contamination by cluster members. These uncertainties
impact the masses at the $5-10\%$ level \btxt{for individual clusters}. We estimated the effect of the lensing calibration uncertainties by examining how the slopes of M-T$_X$ and M-L$_X$ relations change when \boldtext{decreasing the mass of all COSMOS systems by 5 \% while increasing CCCP masses by 5 \% and vice versa}. We find that the effect is small at 3 \% and 5 \% for M-T$_X$ and M-L$_X$ and do not include this effect in the quoted statistical uncertainties.

\begin{figure}
\includegraphics[width=\columnwidth]{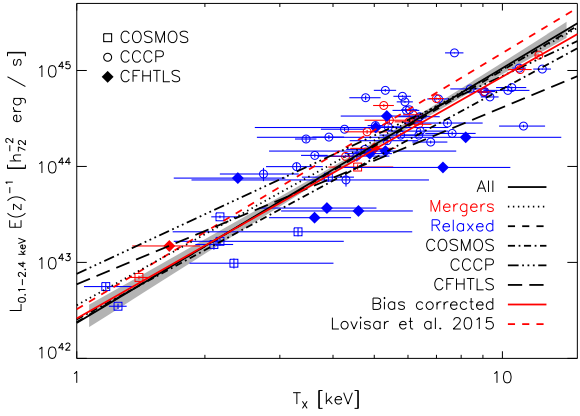}
\caption{The scaling of core-excised X-ray temperature T$_{X}$ to core-excised luminosity L$_X$. The black solid line and grey shaded region shows the best-fit relation and statistical uncertainty fitted to all data, the red solid line shows the corresponding bias corrected relation. The dotted line shows the relation fitted to relaxed clusters (blue data) and dashed line to merging clusters (red data). \boldtext{The dot-dashed and long dashed lines shows relations fitted independently to each survey and the red dashed line is the best-fit uncorrected relation from \citet{Lovisari15}.} Errors on data indicate statistical uncertainties. 
}
\label{fig:ltrel}
\end{figure}

\begin{figure*}
\includegraphics[width=\textwidth]{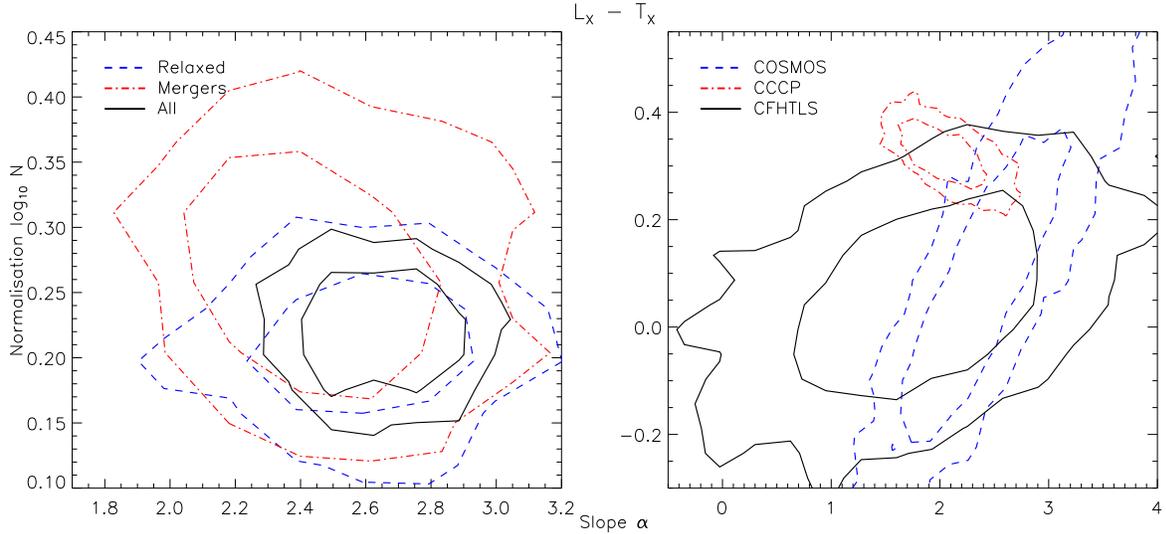}
\caption{\boldtext{Confidence contours for the posterior distributions of slope and normalisation at 68\% and 95\% significance for the  L$_X$-T$_X$ relations fitted to each respective subsample.}
}
\label{fig:ltcont}
\end{figure*}

\subsubsection{Bias correction}
\label{sect:bias}

\boldtext{The \citet{Kelly07} regression method attempts to correct for sampling effects in the independent variable (x-direction). Since we deal with X-ray selected samples of galaxy clusters, we are thus able to correct for possible residual Malmquist bias due to the covariance between the studied parameter and the parameter used to select the clusters by keeping L$_X$ or T$_X$  as the independent variable. However, the regression method determines the scatter only for the dependent variable, and assumes no intrinsic scatter for the independent variable. Consequently, we first have to  determine the scatter in L$_X$ and T$_X$ at fixed mass and add these to the statistical errors. }

\boldtext{Therefore we first measure the global inverted relation with mass as the independent variable to determine the scatter in L$_X$ and T$_X$. We assume that the intrinsic scatter of mass measurement using weak lensing with respect to the true mass is 0.2 in natural logarithm units \citep{Becker11}, and add this value to the mass errors for every fit having mass as the independent variable. As shown by \citet{Vikhlinin09}, the value of the scatter is independent of a possible bias in the slope.   }

\boldtext{The correction term due to  Eddington bias is $$\sigma^2 \ln(10) \frac{d \alpha (\ln(M))}{d ~\ln(M)}$$ \citep{Leauthaud10}, where sigma is the total (statistical and intrinsic) scatter for the parameter in dex, ln(10) is a correction term for using scatter in units of dex and $\alpha$ is a slope of the mass function. We compute the mass-function related term using the parametrisation of \citet{vandenBosch02} and the assumed cosmology. The correction term for mass, L$_X$ and T$_X$ are computed individually for each system in the sample, and we subtract these from the measured values.}

\boldtext{For total scatter in L$_X$ and T$_X$, we use the summed square of the statistical errors and measured intrinsic scatter. The value for the total scatter in weak lensing masses, which correspond to the a convolution of the data quality and the intrinsic scatter, is assumed to be 0.3 in natural logarithm units. This value is used both as the total scatter term for mass and to smooth the theoretical mass function to establishing the derivative of the distribution of clusters as a function of weak lensing mass. Using  weak lensing mass as opposed to the true mass  yields smaller slopes for the mass function. }

\boldtext{We refer to the measurements corrected for Eddington bias and scaling relations fitted to the corrected measurements as bias corrected (BC). The bias correction is discussed in more detail in \citet{Leauthaud10}. Contrary to \citet{Leauthaud10}, who used the global slope of the mass function, we use a local one for each system. While both methods lead to small global changes, using the local slope leads to sizeable corrections in particular for the CCCP sample, which contains a large number of massive clusters at relatively high redshifts. We show the bias corrections for individual systems in Fig. \ref{fig:bias} and list them in Appendix \ref{sec:biastable}.}

\boldtext{As the \citet{Kelly07} fitting routine corrects for Malmquist bias in the independent variable, our bias corrected M-L$_X$ and M-T$_X$ relations are fully corrected for observational biases, whereas there might be some residual covariance affecting the L$_X$-M, T$_X$-M and L$_X$-T$_X$ relations. However, we expect the effect for the global relation to be small. We also explored fits performed individually for each survey (accounting separately for Malmquist bias) and combining the posterior distributions, but found that the combined posterior not to be as constraining as the combined dataset.}

\subsubsection{Morphological classification}
\label{sect:subs}

\begin{figure*}
\includegraphics[width=\textwidth]{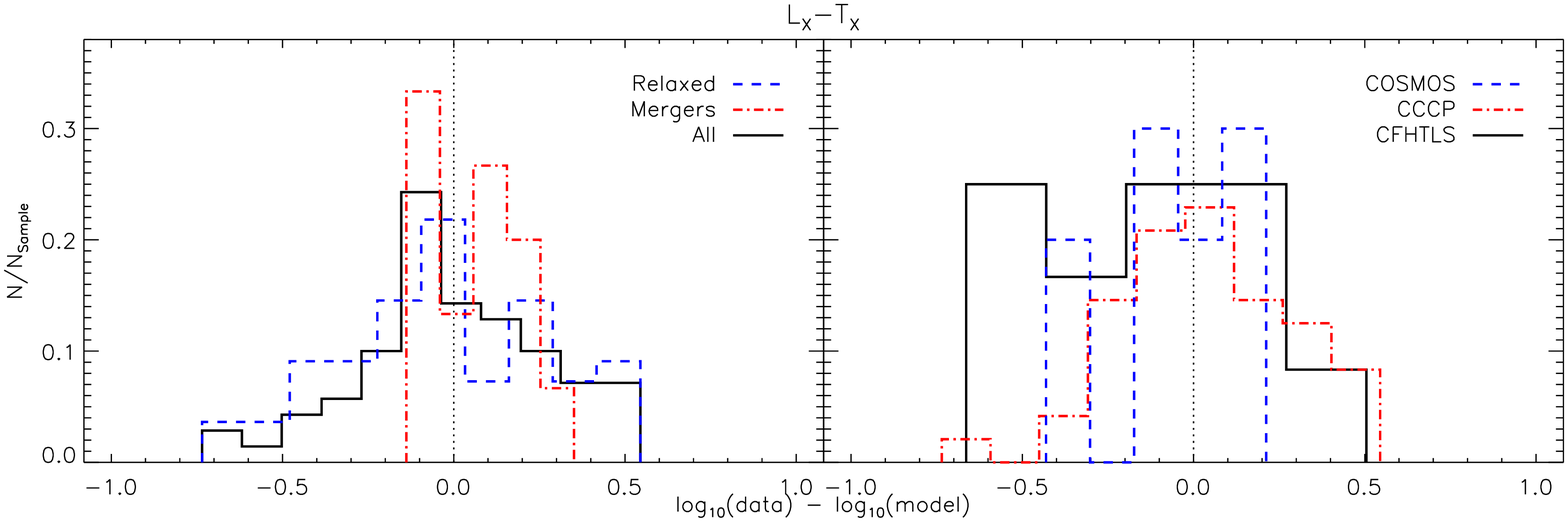}
\caption{\boldtext{The distribution of residuals for each subsample with respect to the L$_X$-T$_X$ relation fitted to the full sample. N$_{\rm Sample}$ is defined as the number of systems in each subsample.}
}
\label{fig:ltres}
\end{figure*}

The distance between the brightest cluster galaxy (BCG) and X-ray surface brightness peak (D$_{\rm BCG}$) has been shown to be a good indicator of the relaxation state by e.g. \citet{Poole07} and \citet{Mahdavi13}. Large values for D$_{\rm BGC}$ indicate significant substructure typical for unrelaxed clusters. We are able to identify BCG locations using the XMM-CFHTLS optical photometry of \citet{Mirkazemi15}. For the XMM-LSS cluster XID102760 we use photometry of \citet{Gozaliasl14}. The location of the X-ray peaks are determined from X-ray photometry presented in this work. For COSMOS and CCCP systems we use D$_{\rm BCG}$ values presented in \citet{Kettula13} and \citet{Mahdavi13} respectively.

We classify clusters with D$_{\rm BGC} <$ 3 \% of R$_{200}$ as relaxed and those with D$_{\rm BGC} \geq$ 3 \% of R$_{200}$ as non-relaxed (which we refer to as mergers or merging clusters). Here R$_{200}$ is the radius inside which the mean density of the cluster corresponds to 200 times the critical density at the redshift of the system. For our sample, 3 \% of R$_{200}$ corresponds to 13 -- 75 kpc and gives \btxt{55} relaxed systems and 15 non-relaxed merging systems (see Fig. \ref{fig:dbcg}). As the CFHTLS and COSMOS samples are selected on X-ray brightness and the CCCP sample, though originally selected on ASCA T$_{X}$, is consistent with well-defined flux based samples \citep{Mahdavi13}, we expect to find a large fraction of relaxed clusters with cool cores associated with high X-ray brightness peaks.

\subsection{L$_{_X}$-T$_{X}$ relation}
\label{sec:lt}

\boldtext{For the L$_{_X}$-T$_{X}$ relation, we adopt L$_0 = 10^{44}$ erg/s and T$_0 =$ 5 keV. The resulting relations and fit parameters are shown in Figs. \ref{fig:ltrel} -- \ref{fig:ltres}, and Table \ref{tab:rel}.}

\boldtext{The scatter in L$_X$ at fixed temperature is $0.15^{+0.04}_{-0.04}$ for the uncorrected relation and $0.10^{+0.04}_{-0.04}$ for the bias corrected relation. The slopes are steeper than the self-similar prediction of 2.0, we get $2.65^{+0.17}_{-0.17}$ in the uncorrected case and $2.52^{+0.17}_{-0.16}$ after bias correction. }

\boldtext{ \citet{Lovisari15} used XMM-Newton observations of a flux-limited set of nearby galaxy groups together with data of the HIFLUGCS clusters from \citet{Hudson10}, resulting in a sample spanning a similar L$_X$ and T$_X$ range as ours. In Fig. \ref{fig:ltrel} we compare their relation corrected for selection bias effects (using full luminosities and core-excised temperatures) to our core-excised relations. We find that their slope is consistent within the uncertainties with our relation, but they predict systematically higher luminosities at fixed temperature because they use total luminosities. }

\begin{figure}
\includegraphics[width=\columnwidth]{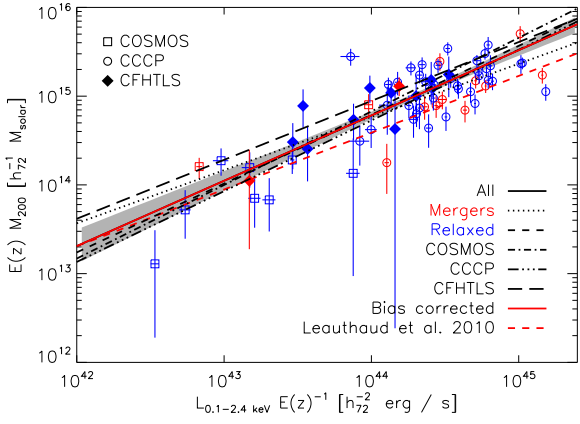}
\caption{\boldtext{The scaling of mass M$_{200}$ to core-excised luminosity L$_X$. The black solid line and grey shaded region shows the best-fit relation and statistical uncertainty fitted to all data, the red solid line shows the corresponding bias corrected relation. The dotted line shows the relation fitted to relaxed clusters (blue data) and dashed line to merging clusters (red data). The dot-dashed and long dashed lines shows relations fitted independently to each survey and the red dashed line is the best-fit uncorrected relation from \citet{Leauthaud10}. Errors on data indicate statistical uncertainties.}
}
\label{fig:mlrel}
\end{figure}

\subsection{M-L$_{X}$ relation}
\label{sec:ml}

X-ray luminosity L$_X$ is the observationally cheapest X-ray observable, requiring only source detection and redshift information for its measurement. Luminosity is hence the mass proxy choice for shallow X-ray surveys, making the mass-luminosity relation potentially a powerful cosmological instrument. 

\begin{figure*}
\includegraphics[width=\textwidth]{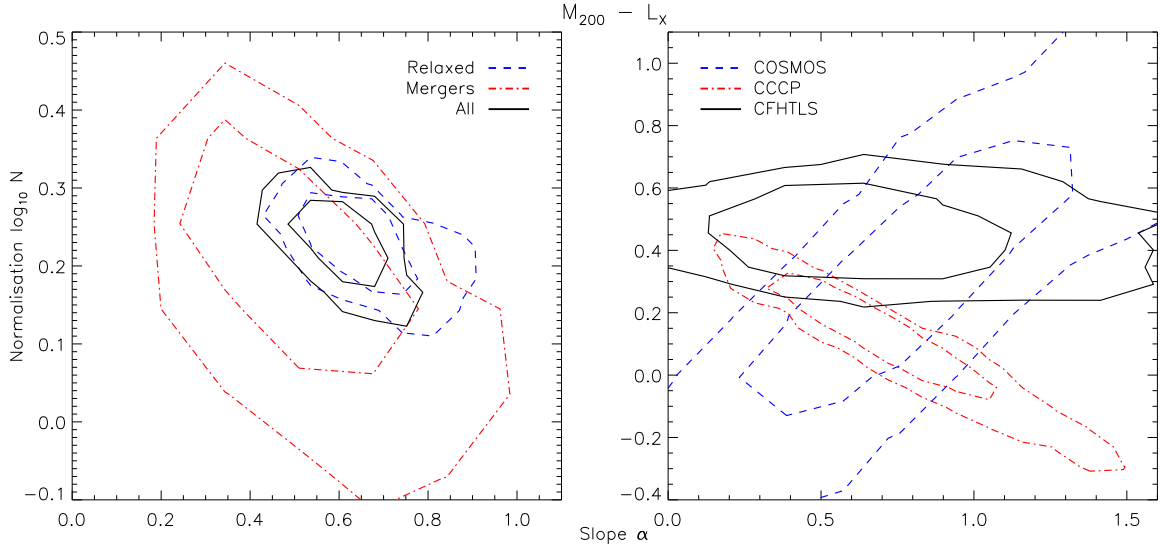}
\caption{\boldtext{Confidence contours for the posterior distributions of slope and normalisation at 68\% and 95\% significance for the  M$_{200}$-L$_X$ relations fitted to each respective subsample.}
}
\label{fig:mlcont}
\end{figure*}

\begin{figure*}
\includegraphics[width=\textwidth]{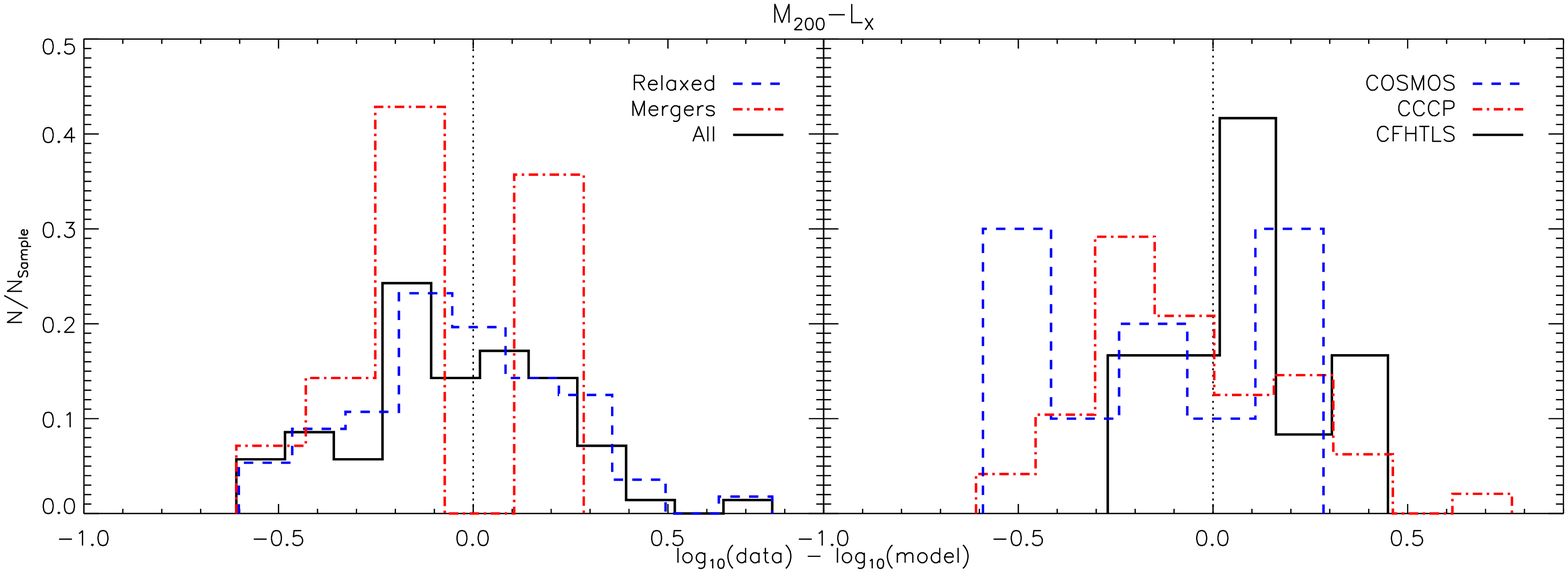}
\caption{\boldtext{he distribution of residuals for each subsample with respect to the M$_{200}$-L$_X$ relation fitted to the full sample. N$_{\rm Sample}$ is defined as the number of systems in each subsample.}
}
\label{fig:mlres}
\end{figure*}

\boldtext{As typically done in the literature, we opt to study the scaling of luminosity to the total mass of the halo given by M$_{200}$, (but also quote the parameters for scaling to M$_{500}$). For the M-L$_X$ relations we set L$_0$ to $10^{44}$ erg/s and M$_0$ to $3 \times 10^{14}$ \msun. The resulting relations and fit parameters are shown in Figs. \ref{fig:mlrel}--\ref{fig:mlres} and Table \ref{tab:rel}. }

\boldtext{The scatter in L$_X$ at fixed mass is $0.33^{+0.03}_{-0.03}$ in the uncorrected case and $0.29^{+0.04}_{-0.03}$ in the  bias corrected case. We obtain a consistent slope for the bias corrected and uncorrected relations, the uncorrected slope is $0.74^{+0.08}_{-0.08}$. The slope is consistent with the purely gravitational self-similar prediction of 0.75. }

\boldtext{Currently the only other M-L$_X$ relation spanning a similar mass range as ours using weak lensing mass calibration is that of \citet{Leauthaud10}. They derived non-core excised luminosities and lensing masses for stacked low-mass galaxy groups in the COSMOS field and combined them with higher mass systems from the literature. Their slope of 0.64 $\pm$ 0.03 is flatter than ours. The \citet{Leauthaud10} relation predicts consistent luminosities with us at low masses, but leading to significant tension at high masses (see Fig. \ref{fig:mlrel}). In addition to the weak lensing measurements, the mass calibration of the low mass \citet{Leauthaud10} sample has been confirmed by magnification analysis \citep{Ford12,Schmidt12} and clustering \citep{Allevato12}. }

\begin{table}
\caption{\boldtext{The fit parameters and intrinsic scatter with the corresponding statistical uncertainties of the scaling relations}.}
\centering
\label{tab:rel}
\begin{tabular}{lccc}
\hline
 & $\alpha$ & $\log_{10} N$ & $\sigma_{\rm \log(A|B)}$ \\ 
\hline
\multicolumn{4}{c}{L$_X$-T$_X$} \\
\hline
All data & $2.65^{+0.17}_{-0.17}$ &  $0.23^{+0.03}_{-0.03}$ & $0.15^{+0.04}_{-0.04}$ \\
Bias corrected & $2.52^{+0.17}_{-0.16}$ &  $0.18^{+0.03}_{-0.03}$ & $0.10^{+0.04}_{-0.04}$ \\
Mergers & $2.46^{+0.27}_{-0.24}$ &  $0.27^{+0.06}_{-0.06}$ & $0.10^{+0.07}_{-0.05}$ \\
Relaxed & $2.62^{+0.22}_{-0.22}$ & $ 0.21^{+0.04}_{-0.04}$ & $0.20^{+0.05}_{-0.04}$ \\
CFHTLS & $1.84^{+0.80}_{-0.76}$ &  $0.06^{+0.12}_{-0.13}$ & $0.34^{+0.13}_{-0.09}$  \\
COSMOS & $2.40^{+0.54}_{-0.46}$ &  $0.08^{+0.21}_{-0.19}$ & $0.17^{+0.12}_{-0.09}$  \\
CCCP & $2.06^{+0.29}_{-0.28}$ &  $0.32^{+0.04}_{-0.04}$ & $0.13^{+0.04}_{-0.04}$  \\
\hline
\multicolumn{4}{c}{L$_X$-M$_{200}$} \\
\hline
All data &  $1.13^{+0.10}_{-0.10}$ & $-0.22^{+0.06}_{-0.06}$ & $0.33^{+0.03}_{-0.03}$ \\
Bias corrected & $1.27^{+0.16}_{-0.15}$ & $-0.38^{+0.06}_{-0.06}$ & $0.29^{+0.04}_{-0.03}$ \\
\hline
\multicolumn{4}{c}{M$_{200}$-L$_X$} \\
\hline
All data & $0.74^{+0.08}_{-0.08}$ &  $0.31^{+0.04}_{-0.04}$ & $0.15^{+0.04}_{-0.04}$ \\
Bias corrected &  $0.74^{+0.09}_{-0.08}$ &  $0.40^{+0.03}_{-0.03}$ & $0.10^{+0.04}_{-0.04}$ \\
Mergers & $0.60^{+0.16}_{-0.15}$ &  $0.29^{+0.10}_{-0.11}$ & $0.21^{+0.10}_{-0.09}$ \\
Relaxed & $0.78^{+0.09}_{-0.09}$ & $ 0.31^{+0.04}_{-0.05}$ & $0.14^{+0.04}_{-0.04}$ \\
CFHTLS & $0.66^{+0.35}_{-0.29}$ &  $0.47^{+0.09}_{-0.10}$ & $0.15^{+0.12}_{-0.08}$  \\
COSMOS & $0.83^{+0.46}_{-0.39}$ &  $0.35^{+0.37}_{-0.34}$ & $0.28^{+0.21}_{-0.13}$  \\
CCCP & $0.80^{+0.38}_{-0.29}$ & $ 0.25^{+0.15}_{-0.21}$ & $0.17^{+0.04}_{-0.05}$  \\
\hline
\multicolumn{4}{c}{M$_{500}$-L$_X$} \\
\hline
All data & $0.70^{+0.08}_{-0.07}$ &  $0.15^{+0.04}_{-0.04}$ & $0.14^{+0.03}_{-0.03}$ \\
\hline
\multicolumn{4}{c}{T$_X$-M$_{500}$} \\
\hline
All data & $0.45^{+0.04}_{-0.04}$ & $-0.02^{+0.02}_{-0.02}$ & $0.11^{+0.01}_{-0.01}$  \\
Bias corrected & $0.48^{+0.06}_{-0.06}$ & $-0.03^{+0.02}_{-0.02}$ & $0.06^{+0.02}_{-0.02}$ \\
\hline
\multicolumn{4}{c}{M$_{500}$-T$_X$} \\
\hline
All data & $1.68^{+0.17}_{-0.17}$ &  $0.08^{+0.03}_{-0.03}$ & $0.14^{+0.03}_{-0.03}$  \\
Bias corrected  & $1.52^{+0.17}_{-0.16}$ & $0.05^{+0.03}_{-0.03}$ & $0.07^{+0.04}_{-0.03}$  \\
Mergers & $1.43^{+0.32}_{-0.31}$ &  $0.05^{+0.07}_{-0.07}$ & $0.18^{+0.09}_{-0.07}$  \\
Relaxed & $1.78^{+0.22}_{-0.21}$ & $ 0.09^{+0.03}_{-0.04}$ & $0.15^{+0.04}_{-0.04}$  \\
CFHTLS & $1.34^{+0.78}_{-0.73}$ &  $0.14^{+0.09}_{-0.10}$ & $0.16^{+0.13}_{-0.08}$  \\
COSMOS & $1.52^{+0.90}_{-0.82}$ & $-0.14^{+0.34}_{-0.34}$ & $0.29^{+0.21}_{-0.14}$  \\
CCCP & $1.18^{+0.31}_{-0.29}$ & $0.14^{+0.05}_{-0.05}$ & $0.17^{+0.03}_{-0.03}$  \\
\hline
\multicolumn{4}{c}{M$_{200}$-T$_X$} \\
\hline
All data & $1.73^{+0.19}_{-0.17}$ &  $0.26^{+0.03}_{-0.03}$ & $0.15^{+0.03}_{-0.03}$  \\
\hline
\end{tabular}

\medskip
$\alpha$ is the slope of the relation, $\log_{10} N$ the normalisation and $\sigma_{\rm \log(A|B)}$ the intrinsic scatter. Bias corrected relations are fitted to the full dataset.
\end{table}

\subsection{M-T$_X$ relation}
\label{sec:mt}

The relation between mass and temperature is the most fundamental among the scaling relations because it provides the physical link between X-ray observations of galaxy clusters and the models of structure formation. If the only source of heating of the gas is gravitational and there is no efficient cooling, the gas temperature is a direct measure of the potential depth, and therefore of the total mass.

\begin{figure}
\includegraphics[width=\columnwidth]{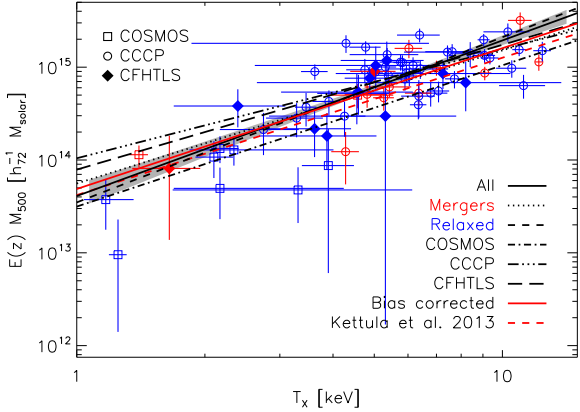}
\caption{\boldtext{The scaling of mass M$_{500}$ to core-excised temperature T$_X$. The black solid line and grey shaded region shows the best-fit relation and statistical uncertainty fitted to all data, the red solid line shows the corresponding bias corrected relation. The dotted line shows the relation fitted to relaxed clusters (blue data) and dashed line to merging clusters (red data). The dot-dashed and long dashed lines shows relations fitted independently to each survey and the red dashed line is the best-fit uncorrected relation from \citet{Kettula13}. Errors on data indicate statistical uncertainties.}
}
\label{fig:mtrel}
\end{figure}

For the M-T$_X$ relation, we opt to study the scaling to M$_{500}$, as is usually done in the literature \boldtext{(but we also quote the parameters of the relation for M$_{200}$)}. The best-fit relations and fit parameters for M$_0$ = $5 \times 10^{14}$ \msun ~and T$_0 = 5.0$ keV are shown in \boldtext{Figs. \ref{fig:mtrel}--\ref{fig:mtres} and Table \ref{tab:rel}. }

\boldtext{We find that T$_X$ is a low-scatter mass proxy, the intrinsic scatter in temperature at fixed mass is $0.11^{+0.01}_{-0.01}$ in the uncorrected case and $0.06^{+0.02}_{-0.02}$ for the fully bias corrected relation. The slope of the uncorrected relation is $1.68^{+0.17}_{-0.17}$. The bias correction results in a slightly shallower slope of $1.52^{+0.17}_{-0.16}$, which is fully consistent with the self-similar prediction of 1.50.}

\boldtext{In Fig. \ref{fig:mtrel} we also compare our relations to the best-fit M--T$_X$ relation from \citet{Kettula13}, where we use CCCP with different temperature measurements as a high-mass sample and five clusters from the 160 Square Degree survey as an intermediate-mass sample to infer a scaling consistent the self-similarity. We find that the best-fit relation of \citet{Kettula13} has a  shallower slope than our uncorrected and bias corrected relations, predicting somewhat lower temperatures for a given mass in the high-mass end.}

\subsection{X-ray cross-calibration}
\label{sect:calib}

\btxt{We investigated the effects of cross-calibration on scaling relations by modifying our XMM based temperatures and luminosities to match Chandra calibration, allowing direct comparison to relations measured with Chandra. We modified our temperatures using the best-fit relations for the full energy band by Eq 3. and Table 2 in \citet{Schellenberger14}. For CFHTLS and COSMOS which are measured with pn only, we used the ACIS--pn relation. For CCCP which uses all three XMM-EPIC detectors (pn, MOS1 and MOS2), we used the values for ACIS--combined XMM. 
}

\btxt{\citet{Nevalainen10} found that Chandra results on average in $\sim$ 2 \% higher fluxes in the soft energy band (0.5--2.0 keV) and $\sim$ 11 \% higher in the hard band (2.0--7.0 keV) than pn. As fluxes are directly related to luminosity, any discrepancy in measured fluxes applies directly to luminosities. \citet{Mahdavi13} reported $\sim$ 3 \% higher  bolometric luminosities for Chandra than for combined XMM. As we measure luminosities in a 0.1--2.4 keV band, we increased our XMM based luminosities by 2 \% in order to match the Chandra calibration.  
}

\btxt{The best-fit parameters of the scaling relations fitted to our modified XMM data are given in Table \ref{tab:cal}, \boldtext{and show the relations in Figs. \ref{fig:ltmantz}--\ref{fig:mtmantz}}. As expected from the small modification to luminosities, we find that modifying luminosities does not affect the resulting relations. However, modifying temperatures drives the slopes of the L$_{X}$-T$_{X}$ and M$_{500}$-T$_{X}$ relations to flatter values. \boldtext{The flattening of the slopes of the  bias corrected L$_{X}$-T$_{X}$ and  M$_{500}$-T$_{X}$ relations are 0.35$\pm$0.16 and 0.23$\pm$0.15 respectively. }  
}

\begin{table}
\caption{\boldtext{The fit parameters and intrinsic scatter with the corresponding statistical uncertainties of the scaling relations with XMM temperatures and luminosities modified to match Chandra calibration.}}
\centering
\label{tab:cal}
\begin{tabular}{lccc}
\hline
 & $\alpha$ & $\log_{10} N$ & $\sigma_{\rm \log(A|B)}$ \\ 
\hline
 \multicolumn{4}{c}{L$_{X}$-T$_X$ Chandra calibration} \\
\hline
All data & $2.25^{+0.15}_{-0.15}$ &  $0.02^{+0.04}_{-0.04}$ & $0.20^{+0.04}_{-0.03}$ \\
Bias corrected & $2.17^{+0.15}_{-0.13}$ & $-0.01^{+0.03}_{-0.03}$ & $0.13^{+0.04}_{-0.04}$ \\
\hline
\multicolumn{4}{c}{M$_{200}$-L$_X$ Chandra calibration} \\
\hline
All data & $0.72^{+0.08}_{-0.07}$ & $0.31^{+0.04}_{-0.04}$ & $0.15^{+0.03}_{-0.03}$ \\
Bias corrected & $1.29^{+0.14}_{-0.13}$ & $-0.07^{+0.03}_{-0.03}$ & $0.08^{+0.04}_{-0.04}$  \\
\hline
 \multicolumn{4}{c}{M$_{500}$-T$_X$ Chandra calibration} \\
\hline
All data & $1.44^{+0.15}_{-0.15}$  &  $-0.05^{+0.04}_{-0.04}$ & $0.16^{+0.03}_{-0.03}$ \\
Bias corrected & $1.29^{+0.14}_{-0.13}$ & $-0.07^{+0.03}_{-0.03}$ & $0.08^{+0.04}_{-0.04}$  \\
\hline
\end{tabular}

\medskip
$\alpha$ is the slope of the relation, $\log_{10} N$ the normalisation and $\sigma_{\rm \log(A|B)}$ the intrinsic scatter. Bias corrected relations are fitted to the full dataset.
\end{table}

\section{Discussion}
\label{sec:disc}

Measurements of a large number of clusters from a wide mass range are needed to gain \btxt{precise} constraints on scaling relations. A large spread in mass improves the constraint on the slope of the scaling and as lensing mass measurements have an intrinsic scatter of $\sim$ 20--30 \% \citep[e.g.][]{Becker11}, several systems in each mass range \boldtext{and a good understanding of systematic uncertainties and observational biases} are needed to \btxt{accurately} recover the average relation.

\begin{figure*}
\includegraphics[width=\textwidth]{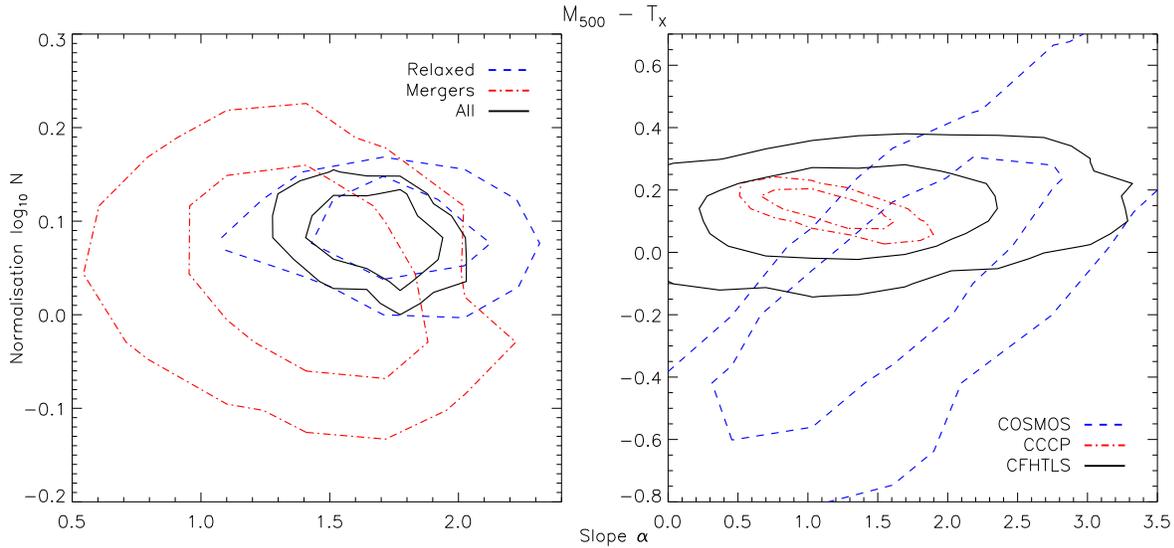}
\caption{\boldtext{Confidence contours for the posterior distributions of slope and normalisation at 68\% and 95\% significance for the  M$_{500}$-T$_X$ relations fitted to each respective subsample.}
}
\label{fig:mtcont}
\end{figure*}

\begin{figure*}
\includegraphics[width=\textwidth]{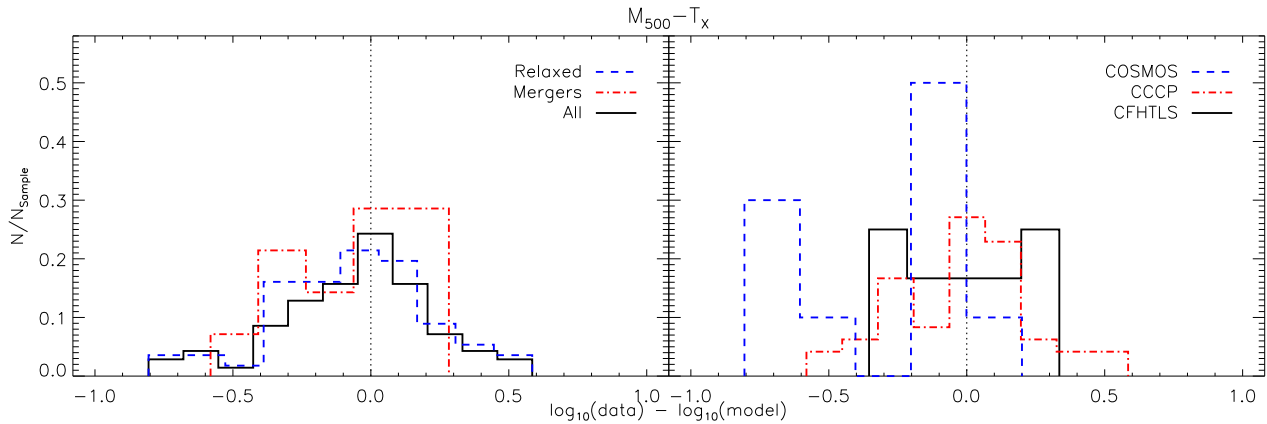}
\caption{\boldtext{The distribution of residuals for each subsample with respect to the M$_{500}$-T$_X$ relation fitted to the full sample. N$_{\rm Sample}$ is defined as the number of systems in each subsample.}
}
\label{fig:mtres}
\end{figure*}

With the inclusion of the 12 low-mass clusters analysed in this work we have more than doubled the number of systems at low and intermediate masses available in the sample used for lensing calibrated scaling relations. Previously the only individual low-mass systems with lensing and X-ray measurements were 10 groups from the COSMOS field, which extend to a larger redshift and thus possibly affected \boldtext{by} evolutionary effects \citep[e.g.][]{Jee11}. On the other hand, the \btxt{there is extensive recent and ongoing observational efforts to obtain mass calibration for massive clusters by e.g. LoCuSS \citep{Okabe10}, CCCP \citep{Mahdavi13} and Weighing the Giants (WtG) \citep{vdl14b}}.

\boldtext{The systems analysed in this work increase the statistical power of the low-mass end and thus improve the \btxt{precision of the} constraint. In addition, we include a correction for Eddington bias. This renders our sample ideal to study mass dependent effects and deviations from self-similar scaling.
}

\subsection{Bias correction} 
\label{sect:biasrel}

\boldtext{As the Eddington bias correction affects the slope of the relation, it is important in order to understand possibly mass dependent deviations from self-similarity. In addition to affecting the slope, the bias correction results in a decrease in scatter. The decreased scatter is an effect of the mass dependence of the bias correction, which drives preferentially upscattered high-mass systems towards the mean relation. As the strength of the bias correction depends on sample selection, it is important to note that the effects of the corrections differ between different surveys.}

\boldtext{As Eddington bias arises as a consequence of intrinsic scatter and an exponential drop in the population, i.e. the high-mass decline of the mass function, it will also affect cluster simulations incorporating a realistic treatment of the intrinsic scatter about the mean relation. Therefore we want to stress the importance of applying the bias correction for simulated cluster populations which are compared to our bias corrected relations.  A full cosmological modelling of cluster core-excised L$_X$ or T$_X$ function should include a convolution of the cluster mass function and bias corrected scaling relation with a log-normal distribution describing the scatter term about the mean relation. }

\subsection{Sensitivity to high-mass sample}
\label{sec:mantz}

\boldtext{In order to test the sensitivity of the global relations to the sample, we replace CCCP with a different high-mass sample. We construct the new sample by correlating the Chandra and ROSAT X-ray measurements of the X-ray selected sample presented in \citet{Mantz10} with the compilation of published weak lensing mass measurements by \citet{Sereno14b}. We find 42 clusters with core-excised temperatures, core-excised soft band X-ray luminosities and weak lensing masses. We refer to this sample as the literature high-mass sample and present the measurements in Appendix \ref{sec:high-m}. The lensing masses are from various sources and consequently suffer from different uncertainties.}

\begin{figure}
\includegraphics[width=\columnwidth]{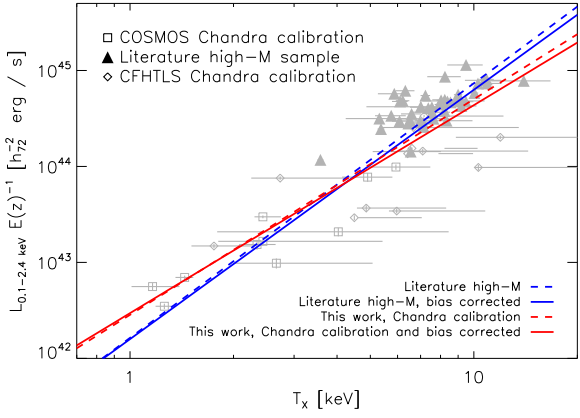}
\caption{\boldtext{Comparison of L$_X$-T$_X$ relations using different high-mass samples, blue lines show relations using the literature sample, red lines using CCCP converted to Chandra calibration. Solid lines show the bias corrected relations, dashed lines the uncorrected lines. The high-mass samples are combined with COSMOS and CFHTLS data converted to Chandra calibration. COSMOS and CFHTLS data converted to Chandra calibration and measurements of the literature high-mass sample are shown in grey.}
}
\label{fig:ltmantz}
\end{figure}

\begin{figure}
\includegraphics[width=\columnwidth]{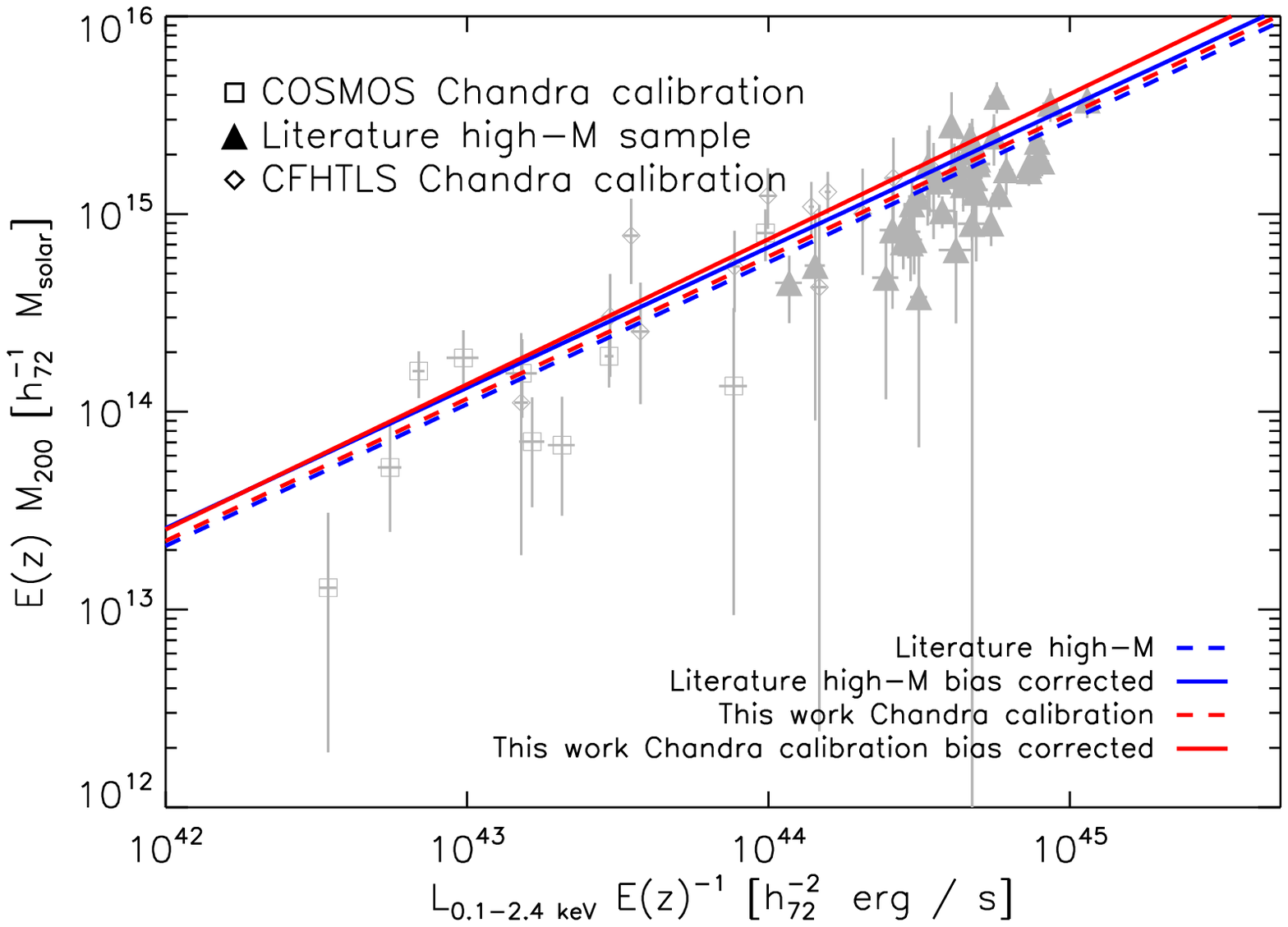}
\caption{\boldtext{Comparison of M-L$_X$ relations using different high-mass samples, blue lines show relations using the literature sample, red lines using CCCP converted to Chandra calibration. Solid lines show the bias corrected relations, dashed lines the uncorrected lines. The high-mass samples are combined with COSMOS and CFHTLS data converted to Chandra calibration. COSMOS and CFHTLS data converted to Chandra calibration and measurements of the literature high-mass sample are shown in grey.}
}
\label{fig:mlmantz}
\end{figure}

\begin{figure}
\includegraphics[width=\columnwidth]{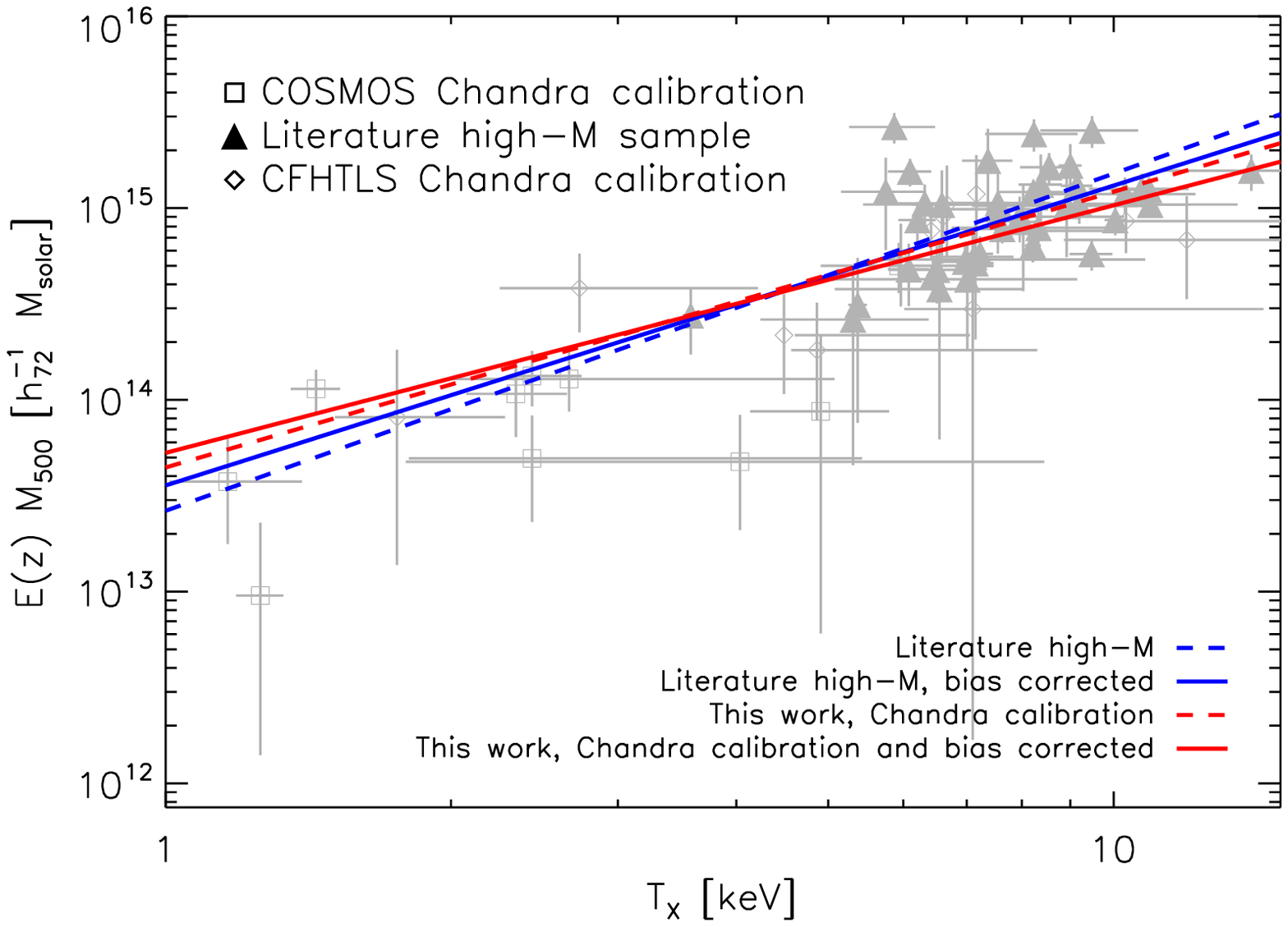}
\caption{\boldtext{Comparison of M-T$_X$ relations using different high-mass samples, blue lines show relations using the literature sample, red lines using CCCP converted to Chandra calibration. Solid lines show the bias corrected relations, dashed lines the uncorrected lines. The high-mass samples are combined with COSMOS and CFHTLS data converted to Chandra calibration. COSMOS and CFHTLS data converted to Chandra calibration and measurements of the literature high-mass sample are shown in grey.}
}
\label{fig:mtmantz}
\end{figure}

\boldtext{As 36 of the clusters in the literature sample have temperatures measured with Chandra and 6 with ASCA, we assume the calibration of the sample to match that of Chandra. We fit L$_X$-T$_X$, M-L$_X$ and M-T$_X$ relations to a sample consisting of the literature sample and COSMOS and CFHTLS data modified to match Chandra calibration (see Table \ref{tab:mantz}). We also apply Eddington bias corrections to this sample and fit bias corrected relations. We show the data and relations and compare them to the corresponding relations using CCCP converted to Chandra calibration as the high mass sample in Figs. \ref{fig:ltmantz}--\ref{fig:mtmantz}. The literature high-mass sample results in systematically steeper relations with lower scatter than CCCP. We also fitted the relations using a subset of the literature sample consisting only of  WtG and CLASH clusters with lensing measurements from \citet{Applegate14} and \citet{Umetsu14}, but found that this had a very small effect.} 

\boldtext{Based on the reported cross-calibration discrepancies, we expect flatter L$_X$-T$_X$ and M-T$_X$ relations for the Chandra based literature sample than for our observed uncorrected XMM-data (as demonstrated in Section \ref{sect:calib}). For M-L$_X$ relation we both expect and find consistent relations, demonstrating consistency in mass and L$_X$ measurements. However, in case of the L$_X$-T$_X$ and M-T$_X$ relations we find that slopes obtained using the literature sample matches the uncorrected XMM-based relations using CCCP, which are steeper than the relations corrected for Chandra calibration. This  demonstrates some tension in the X-ray temperatures of the high mass samples. One possible source of uncertainty is that we use the locally calibrated relation of \citet{Schellenberger14} to convert our XMM based temperatures to match Chandra calibration.} 

\boldtext{Overall, this shows that even after proper accounting for observational biases and considering X-ray cross-calibration issues, differences between samples persist. This variance between samples is still the dominant effect leading to discrepant scaling relations. }

\begin{table}
\caption{\boldtext{The M-L$_X$ relation after replacing CCCP data with the literature sample from \citet{Mantz10} and \citet{Sereno14b} to check the sensitivity of the scaling relations.}}
\centering
\label{tab:mantz}
\begin{tabular}{lccc}
\hline
 & $\alpha$ & $\log_{10} N$ & $\sigma_{\rm \log(A|B)}$ \\ 
\hline
 \multicolumn{4}{c}{L$_X$-T$_X$ literature high-mass sample} \\
\hline
All data & $2.65^{+0.18}_{-0.18}$ &  $0.07^{+0.04}_{-0.04}$ & $0.18^{+0.04}_{-0.04}$ \\ 
Bias corrected & $2.60^{+0.10}_{-0.13}$ & $ 0.02^{+0.03}_{-0.05}$ & $0.09^{+0.05}_{-0.04}$ \\
\hline
 \multicolumn{4}{c}{M$_{200}$-L$_X$ literature high-mass sample} \\
\hline
All data & $0.72^{+0.07}_{-0.06}$ &  $0.28^{+0.04}_{-0.04}$ & $0.08^{+0.04}_{-0.04}$ \\
Bias corrected &  $0.71^{+0.08}_{-0.08}$ &  $0.35^{+0.03}_{-0.03}$ & $0.07^{+0.03}_{-0.03}$ \\ 
\hline
 \multicolumn{4}{c}{M$_{500}$-T$_X$ literature high-mass sample} \\
\hline
All data &  $1.76^{+0.19}_{-0.18}$ & $-0.05^{+0.04}_{-0.04}$ & $ 0.15^{+0.03}_{-0.03}$ \\
Bias corrected & $1.56^{+0.19}_{-0.17}$ & $-0.05^{+0.03}_{-0.03}$ & $0.07^{+0.03}_{-0.03}$ \\
\hline
\end{tabular}
\medskip

$\alpha$ is the slope of the relation, $\log_{10} N$ the normalisation and $\sigma_{\rm \log(A|B)}$ the intrinsic scatter. The relations are fitted to a combination of COSMOS and CFHTLS data corrected to match Chandra calibration and the literature high-mass sample.
\end{table}

\subsection{Mass dependence}
\label{sec:steep}

\boldtext{We fit scaling relations independently to each of the three surveys making up our sample and use them as approximate mass bins in order to attempt to study the mass dependence of the scaling relations. Unfortunately, the statistical uncertainties of the relations fitted to the low-mass COSMOS and intermediate-mass CFHTLS subsamples are large due to the small number of systems and the relatively small mass range. The constraints for the high-mass CCCP subsample are better due to the larger number of systems in the CCCP sample. The relations are described in Figs. \ref{fig:ltrel}-\ref{fig:ltres}, \ref{fig:mlrel}-\ref{fig:mlres} and \ref{fig:mtrel}-\ref{fig:mtres}, and Table \ref{tab:rel}.  We also experimented with CCCP only relations with masses measured assuming the mass-concentration relation of \citet{DM14}  instead of \citet{Duffy08}, but find no difference in the best-fit parameters.}

\boldtext{For COSMOS, we detect a trend for a larger scatter in mass than the higher mass CFHTLS and COSMOS subsamples. For the M-L$_X$ relation, CFHTLS results in higher normalisations than CCCP, whereas the normalisation of the CCCP only L$_X$-T$_X$ is significantly higher than for COSMOS and CFHTLS. As  CCCP selection is based on both L$_X$ and T$_X$, the CCCP only L$_X$-T$_X$ is susceptible to residual scatter affecting the CCCP L$_X$  (see Section \ref{sect:bias}). This could result in the normalisation being biased high. }  

\begin{figure*}
\includegraphics[width=\textwidth]{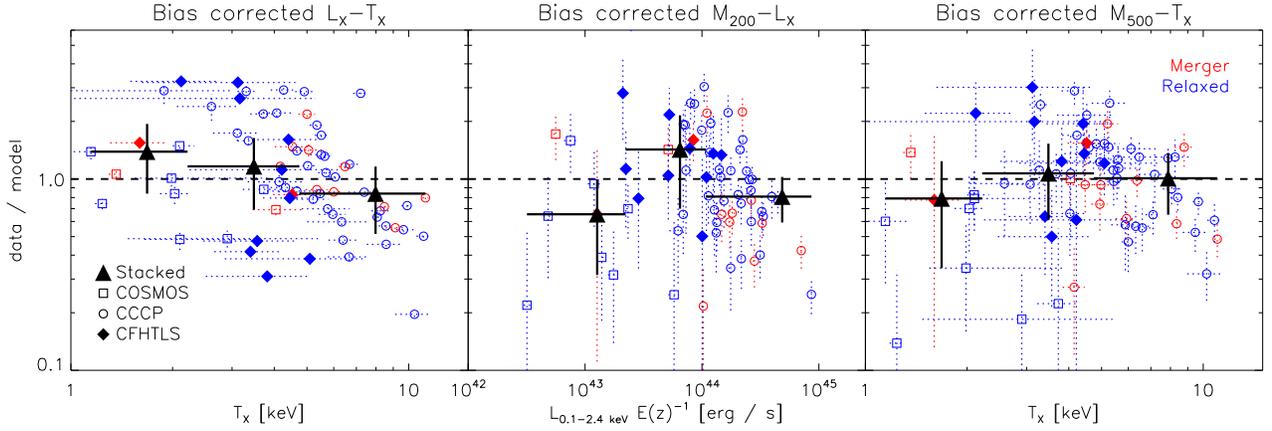}
\caption{\boldtext{Residuals (defined as the ratio of data to model prediction) for the Eddington bias corrected L$_X$-T$_X$ (left panel), M-L$_X$ (middle panel) and M-T$_X$ relations. Blue and red dotted data shows the residuals for individual merging and relaxed systems, squares indicate systems from COSMOS, circles from CCCP and solid diamonds from CFHTLS. Large triangles show the median and median standard deviation of stacked residuals for three mass bins.}
}
\label{fig:rat}
\end{figure*}

\boldtext{We measure residuals (defined as the ratio of data to model prediction) to the  bias corrected relations as a function of luminosity and temperature in Fig. \ref{fig:rat}. We stack the residuals using three mass bins by calculating the median and median standard deviation of the residual in each bin (see Table \ref{tab:rat}). Here a mass dependent deviation from the main relation would drive the median residual away from unity. We use the best-fit relations to determine the luminosity or temperature corresponding to the mass limits of each bin and include the systems falling into the luminosity or temperature range in the stack (here we assume M$_{500} = 0.65$ M$_{200}$). We also repeat the analysis for the bias corrected relation using the literature high-mass sample (Table \ref{tab:rat}). }

\boldtext{For M-L$_X$ and M-T$_X$ relations where we perform full bias corrections, we find consistent behaviour using both data sets, whereas there is tension for the L$_X$-T$_X$ relation. The M-L$_X$ and M-T$_X$ residuals show that low-mass systems (M$_{200} < 2 \times 10^{14}$ \msun) tend to be below the best-fit relation, intermediate  mass systems (M$_{200} = 2-8 \times 10^{14}$ \msun) above the mean relation and high mass systems (M$_{200} > 8 \times 10^{14}$ \msun) above or at the best-fit relation. This is consistent with a mass dependent scaling where low-mass objects follow a steeper scaling than high-mass objects, with the effect being stronger for the mass--luminosity relation than for the mass--temperature relation. This implies that galaxy groups are warmer and more luminous for their mass than clusters. We also see a tendency for steepening at low masses in the L$_X$-T$_X$ relation using CCCP, whereas the literature high-mass sample would result in  opposite behaviour. }

\boldtext{The strong indications of a mass dependence in the M-L$_X$ and M-T$_X$ relations show that there is a need to explore more complicated scaling relation than a single power-law arising from self-similar theory. However, due to the lack of theoretical priors for the functional form and large uncertainties of the data, we do not attempt to model a more complicated scaling. The inferred mass dependence can be attributed to the inclusion of intermediate-mass CFHTLS data and proper accounting for observational biases. Indeed, in \citet{Kettula13} we studied the scaling of lensing mass to temperature of COSMOS groups and clusters from 160SD and CCCP (with different M and T$_X$ measurements than here), and found a single relation connecting groups and high-mass clusters. }


\boldtext{Several previous studies have shown that the scaling relation can deviate from the purely gravitational self-similar prediction and that the deviations become stronger for low-mass systems with masses below a few times 10$^{14}$ \msun \citep[see e.g.][and references therein]{Giodini13}. However, these studies relied on possibly biased HSE mass estimates and this work gives the first indications of different scaling for groups and clusters using accurate lensing masses. }

\boldtext{As shown by Fig. \ref{fig:balance} and e.g. \citet{Giodini10}, the AGN contribution to the energetics of the intracluster gas increases with decreasing mass. As baryonic feedback becomes significant for galaxy groups, energy injection to the intracluster gas in galaxy groups can lead to different scaling for low-mass systems, as indicated in recent simulations by \citet{Planelles14,LeBrun14,Pike14}.
}

\boldtext{Energy injection to the intracluster gas in galaxy groups may also contribute to HSE mass bias in groups. Indeed,  in \citet{Kettula13} we report a HSE mass bias increasing with decreasing mass. This is to be contrasted to the analytical model for non-thermal pressure in galaxy clusters by \citet{Shi14}, who infer a HSE mass bias due to turbulence in the intracluster medium which increases with increasing mass, in line with direct lensing measurements reported in \citet{Mahdavi13}, \citet{vdl14} and \citet{Israel14ii}. However, the model of \citet{Shi14} is contradicted by recent simulations (\citet{Miniati15} and Miniati, F., \textit{private communication}), which show that turbulence scales with the thermal energy and should thus result in a  HSE mass bias which is constant in mass.  As the non-thermal contribution from AGN becomes significant at group levels, the Miniati simulations would thus result in a HSE mass bias consistent with \citet{Kettula13}.} 

\boldtext{Finally, X-ray line emission on group scales may contribute to a break in the mass-to-luminosity relation. Typically the shape of the X-ray spectra of clusters is determined by the bremsstrahlung continuum, but at group masses line emission due to metallicity becomes significant. This results in an extra emission component on top of the bremsstrahlung responsible for $>$ 50 \% of the total X-ray emission, making groups more luminous for their mass. This is not accounted for by the self-similar model and is qualitatively consistent with our findings above.  
}

\begin{table*}
\caption{\boldtext{Stacked residuals of the bias corrected  relations.}}
\centering
\label{tab:rat}
\begin{tabular}{lccc}
\hline
   & M$_{200}$ $< 2 \times 10^{14}$ \msun & M$_{200}$  $= 2-8 \times 10^{14}$ \msun & M$_{200}$ $> 8 \times 10^{14}$ \msun \\
 &  Stacked     &  Stacked      &            Stacked \\
 &  data / model &  data / model &           data / model \\
\hline
L$_{X}$-T$_X$ this work & 1.39 $\pm$ 0.55  & 1.16 $\pm$ 0.47  & 0.84 $\pm$ 0.32 \\
L$_{X}$-T$_X$ literature high-mass sample  & 1.33 $\pm$ 0.53  & 0.73 $\pm$ 0.49  & 0.93 $\pm$ 0.21 \\
\hline
M$_{200}$-L$_X$ this work & 0.65 $\pm$ 0.34 & 1.42 $\pm$ 0.73  & 0.81 $\pm$ 0.22 \\
M$_{200}$-L$_X$ literature high-mass sample  & 0.73 $\pm$ 0.40 & 1.11 $\pm$ 0.45 & 0.93 $\pm$ 0.29 \\ 
\hline
M$_{500}$-T$_X$ this work &  0.79 $\pm$ 0.45  & 1.07 $\pm$ 0.46  & 1.00 $\pm$ 0.36 \\ 
M$_{500}$-T$_X$ literature high-mass sample  & 0.89 $\pm$ 0.11  & 1.12 $\pm$ 0.65  & 0.85 $\pm$ 0.22 \\
\hline
\end{tabular}

\medskip
This work refers to relations combining COSMOS, CFHTLS and CCCP data, literature high-mass sample to relations combining COSMOS and CFHTLS data corrected to match Chandra calibration with the literature high-mass sample.
\end{table*}

\subsection{\btxt{Effects of substructure and triaxiality}}
\label{sec:subst}

\btxt{Simulations by e.g. \citet{Meneghetti10} and \citet{Becker11} indicate that weak lensing masses obtained by fitting NFW profiles to tangential shear profiles suffer from a scatter of $\sim$ 20--25\% \citep[see also discussion in][]{Sereno14}. The main source for the scatter and bias are triaxiality and cluster substructure. Triaxiality and substructure may also bias the resulting masses low by $\sim$ 5\%. As merging clusters per definition display on average stronger deviations from spherical symmetry than relaxed clusters, we expect them to be more strongly affected by scatter and possible bias related to triaxiality and substructure. The large size of the sample allows us to construct subsamples of relaxed and merging clusters to study this effect. \boldtext{We fit relations to the relaxed and merging subsample, and describe them in Figs. \ref{fig:ltrel}-\ref{fig:ltres}, \ref{fig:mlrel}-\ref{fig:mlres} and \ref{fig:mtrel}-\ref{fig:mtres}, and Table \ref{tab:rel}.}
}

\boldtext{In the case of the bias corrected M--L$_X$ and M--T$_X$ relations, which are affected by biases and scatter in lensing masses, we see a trend for a larger scatter in the merging subsample, albeit at a low statistical significance. We do not find any significant differences in the parameters (see Figs. \ref{fig:mlcont} and \ref{fig:mtcont}), but note that the relaxed subsample seems to favour steeper slopes than the merging subsample. This could be evidence for some residual bias originating from the cool core (see Section \ref{sec:xmm}). We also note that possible biases in the slopes do not affect the scatters \citep{Vikhlinin09}.} 

\boldtext{For the L$_X$--T$_X$ relation, which is unaffected by lensing masses, we see the opposite trend in scatter, i.e. mergers have a lower scatter (see Table. \ref{tab:rel}). Once again we find no significant difference in the parameters between merging and relaxed clusters (Fig. \ref{fig:ltcont}), but note that merging clusters might favour a steeper slope and higher normalisation. This is supported by \citet{Bharadwaj14}, who find a steeper slope and higher normalisation for the L$_X$--T$_X$ relation of preferentially relaxed strong cool core groups. However, as \citet{Bharadwaj14} used non core-excised bolometric luminosities, their trend is most likely driven by the inclusion of bright cool cores.}

\begin{figure}
\includegraphics[width=\columnwidth]{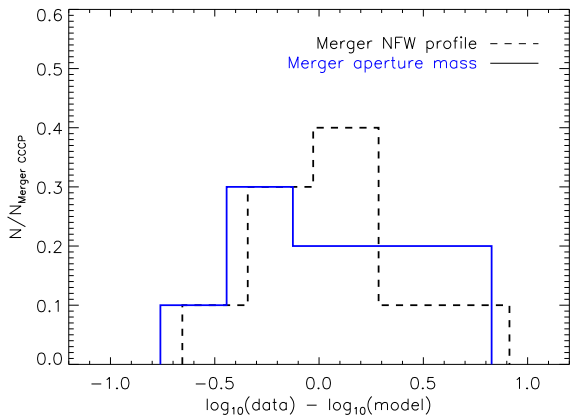}
\caption{\btxt{The mass residuals in the mass--temperature relation for merging CCCP clusters. We show the residuals of M$_{500}$ for all merging CCCP clusters measured using an NFW density profile (dashed black line) and aperture mass (blue solid line) to the best-fit M--T$_X$ relation fitted to all merging clusters in the total sample. }
}
\label{fig:merscat}
\end{figure}

\btxt{We test how strongly the above effects are related to uncertainties arising from assuming an NFW profile by comparing the mass residuals of the M--T$_X$ relation using 11 merging CCCP clusters with mass measurements determined with the NFW assumption and aperture densitometry, available from \citet{Hoekstra15}. Aperture mass relates shear directly to projected density contrast, without any assumptions of geometry. A change in bias would move the residuals systematically to one direction, whereas scatter is determined from the spread of the distribution. We find no difference in scatter or bias using the two mass measurement methods (see Fig. \ref{fig:merscat}).} 

\btxt{Overall, mergers contribute little to the total scatter for X-ray selected samples such as ours. \boldtext{Our measurements also demonstrate that the intrinsic scatter in temperature at fixed mass is significantly lower than in the luminosity at fixed mass.} This shows that temperature is a good low-scatter mass proxy for cluster samples selected on X-ray brightness. However, samples dominated by merging clusters, such as \citet{PlanckXXIX}, might have less scatter using other proxies such as gas mass M$_{gas}$ or ICM thermal energy content Y$_X =$ T$_X \times$ M$_{gas}$. \citet{Mahdavi13} studied these proxies using the high-mass CCCP sample and found that while M$_{gas}$ has lower scatter, Y$_X$ is independent of cluster morphology.
}
 
\section{Summary and conclusions}
\label{sec:sum}

We performed weak lensing and X-ray analysis for a sample of 12 individual low mass clusters within the context of the CFHTLenS and XMM-CFHTLS surveys. This work extends our previous work by inclusion of measurements of intermediate mass systems and provides the first M-L$_X$ relation for low mass systems with individual lensing mass measurements.  We find X-ray luminosities between a few times 10$^{43}$ erg/s and a few times 10$^{45}$ erg/s, temperatures ranging from $\sim$ 2 -- 7 keV and masses M$_{200}$ of  $\sim 10^{14}$ -- 10$^{15}$ \msun. 

Combining the systems analysed in this work with \btxt{lower mass COSMOS and higher mass CCCP} systems from the literature, we end up with a sample of \btxt{70} systems, spanning over two orders of magnitude in mass, three orders of magnitude in luminosity and roughly one order of magnitude in temperature. 

\boldtext{We present a correction for Eddington bias and also apply a Malmquist bias correction for the independent variable (x-direction). As our samples are X-ray selected, we are able to provide fully bias corrected M-L$_X$ and M-T$_X$ relations. By quoting the relations and intrinsic scatters of the parameters, we  provide the current limitations for X-ray luminosity and temperature as cluster mass proxies. We find that the scatter in T$_X$ at fixed mass is significantly lower than that of L$_X$. Though observationally more expensive than L$_X$, this feature renders T$_X$  an attractive mass proxy for use in cosmological work.}

\boldtext{We use the three surveys making up the sample as overlapping mass bins to study mass dependent effects. As the relations fitted to individual surveys suffer from large statistical uncertainties, we do not find any statistically significant effects. Inspecting residuals for the bias corrected relations, we see for the first time indications that galaxy groups are more luminous and warmer for their mass than clusters using accurate lensing masses, implying a steepening in the scaling relations. We expect this steepening to be stronger for luminosity than for temperature. A steepening implies the need for a more complicated scaling than a single power-law predicted from the purely gravitational self-similar model. }

\boldtext{We construct a high-mass sample from the literature to investigate the sensitivity of the relations to the sample. Even after accounting for observational biases and X-ray cross-calibration issues, the literature sample leads to steeper L$_X$-T$_X$ and M-T$_X$ relations, demonstrating that variance between samples is the dominant effect leading to discrepant scaling. However, the inferred mass dependence of the relations is also present with the literature high-mass sample. }

\boldtext{We divide the sample into subsamples of relaxed and merging clusters based on the offset between the X-ray peak and the BCG to investigate the morphology dependence of the scaling. For M--L$_X$ and M--T$_X$ relations which include lensing masses, we find that mergers may result in enhanced scatter, which we attribute to cluster triaxiality and substructure. For the L$_X$--T$_X$ relation which is independent of lensing measurements, we find the opposite trend in scatter.} We study if using aperture mass measurements instead of assuming an NFW profile improves the mass measurements for merging systems, but find no significant effect. For the overall relations fitted to the full sample, we find that mergers contribute little. However, for samples dominated by merging systems, lensing mass calibration using other methods than a single NFW profile \btxt{may lead to improved mass calibration.} 

\btxt{We also explore the effects of X-ray cross-calibration and provide scaling relations with our XMM-Newton based temperatures and luminosities converted to match Chandra calibration. We find that Chandra calibration leads to flatter slopes for   L$_X$--T$_X$ and M--T$_X$ relations, whereas the M--L$_X$ relation is unaffected.}

\boldtext{In conclusion, our work provides a correction for Eddington bias and fully bias corrected scaling relations over a large mass range. We demonstrate the importance of having well understood samples on all mass scales. We detect the first indications of mass dependent scaling relations using weak lensing masses and demonstrate the need for more observations of low mass systems in order to accurately measure the inferred mass dependence. }

\bigskip
Author Contributions: All authors contributed to the development and writing of this paper.  The authorship list reflects the lead authors of this paper (KK, SG, EvU, HH, AF, ML) followed by two alphabetical groups.  The first alphabetical group includes key contributers to the science analysis and interpretation in this paper, the founding core team and those whose long-term significant effort produced the final CFHTLenS and XMM-CFHTLS data products.  The second group covers members of the CFHTLenS and XMM-CFHTLS team who made a significant contribution to the project.  

\section*{Acknowledgments}

\boldtext{The authors thank the anonymous referee for useful comments and suggestions.} This work is based on observations obtained with MegaPrime/MegaCam, a joint project of CFHT and CEA/DAPNIA, at the Canada-France-Hawaii Telescope (CFHT) which is operated by the National Research Council (NRC) of Canada, the Institut National des Sciences de l'Univers of the Centre National de la Recherche Scientifique (CNRS) of France, and the University of Hawaii. This research used the facilities of the Canadian Astronomy Data Centre operated by the National Research Council of Canada with the support of the Canadian Space Agency. CFHTLenS data processing was made possible thanks to significant computing support from the NSERC Research Tools and Instruments grant program, and to HPC specialist Ovidiu Toader.  

KK acknowledges support from Arvid och Greta Ohlins fond and Magnus Ehrnrooth foundation. KK \& AF acknowledge Academy of Finland award, decision 1266918. AF, ML \& MM  have been supported by a DLR project 50 OR 1013 to MPE.   CH, HHo, \& BR acknowledge support from the European Research Council under the EC FP7 grant numbers 240185 (CH), 279396 (HHo),  \& 240672 (BR).   TDK acknowledges support from a Royal Society University Research Fellowship.  TE is supported by the Deutsche Forschungsgemeinschaft through project ER 327/3-1 and is supported by the Transregional Collaborative Research Centre TR 33 - "The Dark Universe". HHi is supported by the Marie Curie IOF 252760, a CITA National Fellowship, and the DFG grant Hi 1495/2-1. HHo, SG acknowledge support from the Netherlands Organisation for Scientiﬁc Research grant number 639.042.814.    YM acknowledges support from CNRS/INSU (Institut National des Sciences de l'Univers) and the Programme National Galaxies et Cosmologie (PNCG).  LVW acknowledges support from the Natural Sciences and Engineering Research Council of Canada (NSERC) and the Canadian Institute for Advanced Research (CIfAR, Cosmology and Gravity program). JPK acknowledge generous support from the ERC advanced grant: LIDA. LF acknowledges support from NSFC grants 11103012 \& 11333001 \& Shanghai Research grant 13JC1404400. MJH acknowledges support from the Natural Sciences and Engineering Research Council of Canada (NSERC).

\appendix

\section{COSMOS and CCCP luminosities}
\label{sec:lum}

\boldtext{The core-excised soft band luminosities for COSMOS are given in Table \ref{tab:cosmos} and for CCCP in Table \ref{tab:cccp}.}

\begin{table}
\caption{\btxt{The \boldtext{core-excised soft band} X-ray luminosities of the COSMOS systems.}}
\centering
\label{tab:cosmos}
\begin{tabular}{lc}
\hline
COSMOS & L$_X$  \\ 
xid    & 10$^{43}$ erg sec$^{-1}$ \\
\hline
  11  & 3.24 $\pm$ 0.11 \\
  17  & 1.81 $\pm$ 0.21 \\
  25  & 0.36 $\pm$ 0.02 \\
  29  & 1.14 $\pm$ 0.14 \\
  120 & 12.02 $\pm$ 1.27 \\
  149 & 0.72 $\pm$ 0.03 \\
  193 & 0.61 $\pm$ 0.05 \\
  220 & 14.38 $\pm$ 0.93 \\
  237 & 1.93 $\pm$ 0.18 \\
  262 & 2.42 $\pm$ 0.25 \\
\hline
\end{tabular}
\end{table}

\begin{table}
\caption{\boldtext{The core-excised soft band X-ray luminosities of the CCCP systems.}}
\centering
\label{tab:cccp}
\begin{tabular}{lc}
\hline
Cluster & L$_X$  \\ 
name   & 10$^{43}$ erg sec$^{-1}$ \\
\hline
  3C295            & 19.24    $\pm$        0.79 \\       
  Abell0068        & 43.66    $\pm$        1.63  \\          
  Abell0115N       & 35.35    $\pm$       0.70   \\        
  Abell0115S       & 47.58    $\pm$        1.52 \\
  Abell0209        & 55.94    $\pm$       0.81 \\
  Abell0222        & 22.49    $\pm$        1.16  \\          
  Abell0223S       & 19.90    $\pm$        0.69 \\          
  Abell0267        & 30.85    $\pm$        0.81 \\          
  Abell0370        & 40.13    $\pm$        1.39  \\
  Abell0383        & 21.17    $\pm$        1.60  \\          
  Abell0520        & 56.26    $\pm$        1.22  \\          
  Abell0521        & 53.48    $\pm$        1.20 \\
  Abell0586        & 26.57    $\pm$        1.16 \\
  Abell0611        & 30.51    $\pm$        0.95  \\         
  Abell0697        & 76.88    $\pm$         1.80 \\
  Abell0851        & 36.15    $\pm$        1.034\\
  Abell0959        & 21.45     $\pm$       1.72   \\         
  Abell0963        & 43.14    $\pm$        1.13  \\          
  Abell1689        & 64.57    $\pm$        0.45  \\
  Abell1763        & 60.15    $\pm$        1.42  \\          
  Abell1835        & 68.57    $\pm$        0.52  \\        
  Abell1914        & 64.83    $\pm$        0.91  \\
  Abell1942        & 14.57    $\pm$        0.70  \\         
  Abell2104        & 66.88    $\pm$        2.11  \\          
  Abell2111        & 33.98    $\pm$        2.50 \\
  Abell2163        & 159.92   $\pm$        2.55 \\           
  Abell2204        & 57.07    $\pm$       0.42   \\        
  Abell2218        & 37.63    $\pm$       0.46 \\
  Abell2219        & 170.81   $\pm$       1.66  \\          
  Abell2259        & 24.87    $\pm$       1.17  \\          
  Abell2261        & 58.15    $\pm$       3.22   \\         
  Abell2390        & 118.57   $\pm$        1.73\\
  Abell2537        & 32.93    $\pm$        1.23  \\          
  CL0024.0+1652    & 8.87     $\pm$       1.30 \\
  MACSJ0717.5+3745 & 137.73   $\pm$        2.31  \\          
  MACSJ0913.7+4056 & 26.79    $\pm$        0.69   \\       
  MS0015.9+1609    & 83.46    $\pm$        2.10   \\        
  MS0440.5+0204    & 9.15     $\pm$       1.44    \\        
  MS0451.6-0305    & 86.60    $\pm$        2.66  \\           
  MS0906.5+1110    & 28.08    $\pm$        0.97  \\         
  MS1008.1-1224    & 24.74    $\pm$        1.22 \\
  MS1231.3+1542    & 14.30    $\pm$        0.49  \\         
  MS1358.1+6245    & 27.67    $\pm$        1.60  \\          
  MS1455.0+2232    & 30.35    $\pm$        0.90  \\         
  MS1512.4+3647    & 12.10    $\pm$        1.10  \\          
  MS1621.5+2640    & 27.71    $\pm$        1.27 \\
  RXJ1347.5-1145   & 131.61   $\pm$         2.01\\
  RXJ1524.6+0957   & 16.93    $\pm$        2.03  \\  
\hline
\end{tabular}
\end{table}

\section{Eddington bias corrections}
\label{sec:biastable}

\boldtext{The Eddington bias corrections for CFHTLS, COSMOS and CCCP systems are given in Tables \ref{tab:cfhtlsbias}, \ref{tab:cosmosbias} and \ref{tab:cccpbias} respectively.}

\begin{table*}
\caption{\boldtext{The Eddington bias corrections for CFHTLS systems.}}
\centering
\label{tab:cfhtlsbias}
\begin{tabular}{lcccc}
\hline
xid & $ \frac{d \alpha(ln(M)}{d ln(M)}$ & M$_{rat}$ & T$_{X, rat}$ & L$_{X, rat}$ \\ 
\hline
110090 & 0.988 &       0.915 &       0.939	 &       0.763 \\
110460 & 2.265 &       0.816 &       0.703	 &       0.538 \\
110850 & 1.410 &       0.881 &       0.887	 &       0.681 \\
110860 & 0.925 &       0.920 &       0.920	 &       0.776 \\
111180 & 2.220 &       0.819 &       0.903	 &       0.547 \\
210010 & 2.019 &       0.834 &       0.910	 &       0.577 \\
210020 & 0.576 &       0.949 &       0.968	 &       0.854 \\
210630 & 1.334 &       0.887 &       0.792	 &       0.694 \\
210740 & 1.792 &       0.851 &       0.830	 &       0.613 \\
210910 & 2.610 &       0.791 &       0.619	 &       0.490 \\
210970 & 3.042 &       0.761 &       0.824	 &       0.437 \\
102760 & 2.247 &       0.817 &       0.385	 &       0.542 \\
\hline
\end{tabular}

\medskip
$ \frac{d \alpha(ln(M)}{d ln(M)}$ is the slope of the mass function, M$_{rat}$, T$_{X, rat}$ and L$_{X, rat}$ are the ratio of the Eddington bias corrected mass, temperature and luminosity to the uncorrected values.
\end{table*}

\begin{table*}
\caption{\boldtext{The Eddington bias corrections for COSMOS systems.}}
\centering
\label{tab:cosmosbias}
\begin{tabular}{lcccc}
\hline
xid & $ \frac{d \alpha(ln(M)}{d ln(M)}$ & M$_{rat}$ & T$_{X, rat}$ & L$_{X, rat}$ \\ 
\hline
11  & 0.806 &       0.930 &       0.967	 &       0.803 \\
17  & 0.797 &       0.931 &       0.966	 &       0.800 \\
25  & 0.235 &       0.979 &       0.990	 &       0.937 \\
29  & 0.857 &       0.926 &       0.898	 &       0.787 \\
120 & 0.959 &       0.917 &       0.954	 &       0.766 \\
149 & 0.699 &       0.939 &       0.972	 &       0.826 \\
193 & 0.436 &       0.961 &       0.979	 &       0.8863 \\
220 & 2.274 &       0.815 &       0.884	 &       0.536 \\
237 & 0.538 &       0.952 &       0.914	 &       0.861 \\
262 & 0.526 &       0.953 &       0.877	 &       0.864 \\
\hline
\end{tabular}

\medskip
$ \frac{d \alpha(ln(M)}{d ln(M)}$ is the slope of the mass function, M$_{rat}$, T$_{X, rat}$ and L$_{X, rat}$ are the ratio of the Eddington bias corrected mass, temperature and luminosity to the uncorrected values.
\end{table*}

\begin{table*}
\caption{\boldtext{The Eddington bias corrections for CCCP systems.}}
\centering
\label{tab:cccpbias}
\begin{tabular}{lcccc}
\hline
Name & $ \frac{d \alpha(ln(M)}{d ln(M)}$ & M$_{rat}$ & T$_{X, rat}$ & L$_{X, rat}$ \\ 
\hline
3C295	 & 2.7246   	 & 0.783 & 0.890	 & 0.476 \\
A68	     & 2.3183   	 & 0.812 & 0.910	 & 0.532 \\
A115N	 & 1.4342   	 & 0.879 & 0.945	 & 0.677 \\
A115S	 & 1.5806   	 & 0.868 & 0.939	 & 0.651 \\
A209	 & 1.7475   	 & 0.855 & 0.933	 & 0.622 \\
A222	 & 1.5154   	 & 0.873 & 0.939	 & 0.661 \\
A223S	 & 1.6967   	 & 0.859 & 0.930	 & 0.630 \\
A267	 & 1.719    	 & 0.857 & 0.932	 & 0.627 \\
A370	 & 4.2331   	 & 0.684 & 0.827	 & 0.316 \\
A383	 & 1.3782   	 & 0.884 & 0.947	 & 0.685 \\
A520	 & 2.423    	 & 0.805 & 0.910	 & 0.518 \\
A521	 & 2.1272   	 & 0.826 & 0.920	 & 0.561 \\
A586	 & 1.2077   	 & 0.897 & 0.947	 & 0.720 \\
A611	 & 1.977    	 & 0.838 & 0.913	 & 0.584 \\
A697	 & 2.5279   	 & 0.797 & 0.900	 & 0.503 \\
A851	 & 3.5767   	 & 0.726 & 0.864	 & 0.378 \\
A959	 & 3.0663   	 & 0.760 & 0.843	 & 0.431 \\
A963	 & 2.1625   	 & 0.824 & 0.919	 & 0.556 \\
A1689	 & 3.5814   	 & 0.725 & 0.871	 & 0.378 \\
A1763	 & 2.6444   	 & 0.789 & 0.901	 & 0.487 \\
A1835	 & 2.9783   	 & 0.766 & 0.888	 & 0.445 \\
A1914	 & 2.2112   	 & 0.820 & 0.918	 & 0.549 \\
A1942	 & 2.3255   	 & 0.812 & 0.910	 & 0.530 \\
A2104	 & 2.3853   	 & 0.808 & 0.909	 & 0.523 \\
A2111	 & 1.8756   	 & 0.845 & 0.917	 & 0.598 \\
A2163	 & 2.6381   	 & 0.789 & 0.904	 & 0.488 \\
A2204	 & 2.7131   	 & 0.784 & 0.899	 & 0.479 \\
A2218	 & 2.4756   	 & 0.801 & 0.907	 & 0.511 \\
A2219	 & 2.0995   	 & 0.829 & 0.921	 & 0.566 \\
A2259	 & 1.6281   	 & 0.864 & 0.930	 & 0.642 \\
A2261	 & 3.2946   	 & 0.744 & 0.871	 & 0.407 \\
A2390	 & 3.1878   	 & 0.752 & 0.884	 & 0.421 \\
A2537	 & 3.1944   	 & 0.751 & 0.823	 & 0.419 \\
CL0024	 & 3.8042   	 & 0.711 & 0.367	 & 0.341 \\
MACS0717 & 5.6275   	 & 0.604 & 0.798	 & 0.217 \\
CL0910	 & 1.8656   	 & 0.846 & 0.924	 & 0.602 \\
MS0016	 & 4.8253   	 & 0.649 & 0.801	 & 0.269 \\
MS0440	 & 1.0189   	 & 0.913 & 0.940	 & 0.749 \\
MS0451	 & 3.6064   	 & 0.724 & 0.855	 & 0.375 \\
MS0906	 & 2.1382   	 & 0.826 & 0.917	 & 0.559 \\
MS1008	 & 2.7792   	 & 0.780 & 0.885	 & 0.469 \\
MS1231	 & 0.78512  	 & 0.932 & 0.968	 & 0.808 \\
MS1358	 & 2.5624   	 & 0.795 & 0.876	 & 0.497 \\
MS1455	 & 2.355    	 & 0.810 & 0.912	 & 0.527 \\
MS1512	 & 1.3264   	 & 0.888 & 0.936	 & 0.694 \\
MS1621	 & 3.4511   	 & 0.734 & 0.849	 & 0.390 \\
RXJ1347	 & 3.5424   	 & 0.728 & 0.870	 & 0.382 \\
RXJ1524	 & 2.0217   	 & 0.834 & 0.890	 & 0.569 \\
\hline
\end{tabular}

\medskip
$ \frac{d \alpha(ln(M)}{d ln(M)}$ is the slope of the mass function, M$_{rat}$, T$_{X, rat}$ and L$_{X, rat}$ are the ratio of the Eddington bias corrected mass, temperature and luminosity to the uncorrected values.
\end{table*}

\section{Literature high-mass sample}
\label{sec:high-m}

\boldtext{We give the X-ray luminosity and temperature measurements and lensing masses of the literature high-mass sample in Table \ref{tab:lit}. The Eddington bias corrections are described in Table \ref{tab:litbias}.}

\begin{table*}
\caption{\boldtext{The literature high-mass sample from \citet{Mantz10} and \citet{Sereno14b}.}}
\centering
\label{tab:lit}
\begin{tabular}{lccccccc}
\hline
Cluster &  $z$ & L$_X$                   & T$_X$ & kT     & $M_\mathrm{500}$ & $M_\mathrm{200}$  & Author \\
name     &       &10$^{43}$ erg sec$^{-1}$ & keV & ref.  & 10$^{14}$ \msun  & 10$^{14}$ \msun   &  code \\
\hline
Abell2029 & 0.0779 & 41.4 $\pm$ 3.9 & 8.22 $\pm$ 0.16 & 2 & 6.501 $\pm$ 1.189 & 10.278 $\pm$ 1.88 & cypriano+04 \\
Abell478 & 0.0881 & 48.8 $\pm$ 4.7 & 7.96 $\pm$ 0.27 & 2 & 9.168 $\pm$ 2.452 & 13.857 $\pm$ 3.707 & okabe+14b \\
Abell2142 & 0.0904 & 64.3 $\pm$ 3.5 & 10.04 $\pm$ 0.26 & 2 & 8.777 $\pm$ 1.476 & 12.457 $\pm$ 2.095 & umetsu+09 \\
Abell2244 & 0.0989 & 27.2 $\pm$ 2.7 & 5.37 $\pm$ 0.12 & 2 & 3.157 $\pm$ 2.391 & 4.678 $\pm$ 3.543 & kubo+09 \\
Abell2034 & 0.113 & 28.8 $\pm$ 2.7 & 7.15 $\pm$ 0.32 & 1 & 5.169 $\pm$ 3.1 & 8.086 $\pm$ 4.849 & okabe\&08 \\
Abell2204 & 0.1511 & 53 $\pm$ 5.2 & 8.55 $\pm$ 0.58 & 2 & 16.051 $\pm$ 2.963 & 23.197 $\pm$ 4.283 & applegate+14 \\
Abell2218 & 0.171 & 33.5 $\pm$ 3.2 & 6.97 $\pm$ 0.37 & 1 & 5.108 $\pm$ 1.358 & 7.697 $\pm$ 2.047 & mahdavi+13 \\
Abell1914 & 0.1712 & 54.4 $\pm$ 5.5 & 9.48 $\pm$ 0.49 & 1 & 5.6 $\pm$ 1.009 & 8.451 $\pm$ 1.523 & mahdavi+13 \\
Abell665 & 0.1818 & 56.5 $\pm$ 5.2 & 8.03 $\pm$ 0.24 & 1 & 8.186 $\pm$ 4.621 & 12.461 $\pm$ 7.035 & pedersen\&07 \\
Abell520 & 0.203 & 64.1 $\pm$ 2 & 7.23 $\pm$ 0.23 & 3 & 5.516 $\pm$ 1.272 & 8.343 $\pm$ 1.925 & mahdavi+13 \\
Abell963 & 0.206 & 34.7 $\pm$ 1.5 & 6.08 $\pm$ 0.3 & 3 & 4.583 $\pm$ 1.637 & 6.623 $\pm$ 2.365 & applegate+14 \\
Abell1423 & 0.213 & 39.7 $\pm$ 2.4 & 5.75 $\pm$ 0.59 & 3 & 11.568 $\pm$ 5.823 & 16.282 $\pm$ 8.196 & dahle06 \\
Abell773 & 0.217 & 47.7 $\pm$ 1.5 & 7.37 $\pm$ 0.45 & 3 & 16.757 $\pm$ 7.814 & 25.985 $\pm$ 12.118 & pedersen\&07 \\
Abell2261 & 0.224 & 56.3 $\pm$ 2.3 & 6.1 $\pm$ 0.32 & 3 & 14.663 $\pm$ 2.394 & 21.246 $\pm$ 4.09 & umetsu+14 \\
Abell1682 & 0.226 & 49.6 $\pm$ 6.1 & 7.01 $\pm$ 2.14 & 3 & 4.014 $\pm$ 2.307 & 6.048 $\pm$ 3.476 & pedersen\&07 \\
Abell1763 & 0.2279 & 72.9 $\pm$ 3.9 & 6.32 $\pm$ 0.4 & 3 & 9.989 $\pm$ 2.516 & 15.329 $\pm$ 3.86 & mahdavi+13 \\
Abell2219 & 0.2281 & 95.8 $\pm$ 5.3 & 10.9 $\pm$ 0.53 & 3 & 11.729 $\pm$ 1.852 & 16.951 $\pm$ 2.677 & applegate+14 \\
Abell2111 & 0.229 & 36.1 $\pm$ 2.2 & 6.51 $\pm$ 0.72 & 3 & 4.498 $\pm$ 1.491 & 6.795 $\pm$ 2.251 & mahdavi+13 \\
Abell267 & 0.23 & 33.6 $\pm$ 1.6 & 7.13 $\pm$ 0.71 & 3 & 5.245 $\pm$ 1.523 & 7.948 $\pm$ 2.308 & mahdavi+13 \\
Abell2390 & 0.2329 & 86.9 $\pm$ 2.9 & 10.28 $\pm$ 0.38 & 3 & 11.183 $\pm$ 2.396 & 16.162 $\pm$ 3.463 & applegate+14 \\
Abell1835 & 0.2528 & 67.3 $\pm$ 2.3 & 9.0 $\pm$ 0.25 & 3 & 15.51 $\pm$ 4.503 & 22.417 $\pm$ 6.508 & applegate+14 \\
Abell68 & 0.2546 & 44.2 $\pm$ 2.7 & 7.56 $\pm$ 0.97 & 3 & 9.171 $\pm$ 1.587 & 13.254 $\pm$ 2.294 & applegate+14 \\
Abell697 & 0.282 & 89.5 $\pm$ 5 & 10.93 $\pm$ 1.11 & 3 & 9.531 $\pm$ 1.303 & 14.694 $\pm$ 2.009 & mahdavi+13 \\
Abell781 & 0.2984 & 51 $\pm$ 3.2 & 7.55 $\pm$ 1.03 & 3 & 9.655 $\pm$ 4.393 & 13.78 $\pm$ 6.27 & dahle06 \\
Abell85 & 0.0557 & 30.4 $\pm$ 2 & 6.45 $\pm$ 0.1 & 2 & 4.579 $\pm$ 1.245 & 7.24 $\pm$ 1.968 & cypriano+04 \\
Abell2597 & 0.0852 & 12.9 $\pm$ 1.3 & 3.58 $\pm$ 0.07 & 1 & 2.803 $\pm$ 1.047 & 4.432 $\pm$ 1.656 & cypriano+04 \\
Abell1689 & 0.1832 & 57.2 $\pm$ 5.7 & 9.15 $\pm$ 0.35 & 1 & 12.614 $\pm$ 1.671 & 16.843 $\pm$ 2.429 & umetsu+11 \\
Abell209 & 0.206 & 58 $\pm$ 2.2 & 8.23 $\pm$ 0.66 & 3 & 11.573 $\pm$ 1.796 & 17.559 $\pm$ 2.993 & umetsu+14 \\
Abell521 & 0.2475 & 58 $\pm$ 2.1 & 6.21 $\pm$ 0.28 & 3 & 8.082 $\pm$ 1.94 & 11.68 $\pm$ 2.803 & applegate+14 \\
Abell2537 & 0.2966 & 38.7 $\pm$ 2.9 & 7.63 $\pm$ 0.86 & 3 & 7.068 $\pm$ 1.113 & 10.841 $\pm$ 1.707 & mahdavi+13 \\
MACSJ1115.8+0129 & 0.355 & 54.3 $\pm$ 2.5 & 9.2 $\pm$ 0.98 & 3 & 9.259 $\pm$ 1.991 & 15.531 $\pm$ 3.385 & umetsu+14 \\
MACSJ0949.8+1708 & 0.384 & 62.3 $\pm$ 4.1 & 8.92 $\pm$ 1.83 & 3 & 8.874 $\pm$ 4.075 & 12.825 $\pm$ 5.889 & applegate+14 \\
MACSJ1731.6+2252 & 0.389 & 74.2 $\pm$ 4.3 & 5.87 $\pm$ 0.61 & 3 & 22.817 $\pm$ 4.087 & 32.977 $\pm$ 5.906 & applegate+14 \\
MACSJ2211.7-0349 & 0.396 & 101.5 $\pm$ 6.3 & 13.97 $\pm$ 2.74 & 3 & 13.447 $\pm$ 2.881 & 19.434 $\pm$ 4.164 & applegate+14 \\
MACSJ0429.6-0253 & 0.399 & 39.1 $\pm$ 2.5 & 8.33 $\pm$ 1.58 & 3 & 6.765 $\pm$ 1.89 & 9.351 $\pm$ 2.984 & umetsu+14 \\
MACSJ1206.2-0847 & 0.439 & 105.5 $\pm$ 6.4 & 10.71 $\pm$ 1.29 & 3 & 10.542 $\pm$ 2.089 & 15.813 $\pm$ 3.58 & umetsu+14 \\
MACSJ0417.5-1154 & 0.443 & 152.9 $\pm$ 9.4 & 9.49 $\pm$ 1.12 & 3 & 21.176 $\pm$ 3.97 & 30.605 $\pm$ 5.738 & applegate+14 \\
MACSJ2243.3-0935 & 0.447 & 115.6 $\pm$ 6.7 & 8.24 $\pm$ 0.92 & 3 & 20.294 $\pm$ 3.865 & 29.33 $\pm$ 5.587 & applegate+14 \\
RXJ0439.0+0715 & 0.2443 & 42.2 $\pm$ 1.6 & 6.59 $\pm$ 0.45 & 3 & 9.753 $\pm$ 4.955 & 13.792 $\pm$ 7.006 & dahle06 \\
Zwicky5247 & 0.229 & 37.4 $\pm$ 2.4 & 5.31 $\pm$ 1.07 & 3 & 2.472 $\pm$ 2.042 & 3.49 $\pm$ 2.883 & dahle06 \\
Zwicky2089 & 0.2347 & 17 $\pm$ 1.3 & 6.55 $\pm$ 1.47 & 3 & 3.55 $\pm$ 2.965 & 5.02 $\pm$ 4.193 & dahle06 \\
Zwicky3146 & 0.2906 & 58.2 $\pm$ 2.6 & 8.38 $\pm$ 0.44 & 3 & 12.071 $\pm$ 5.271 & 18.72 $\pm$ 8.175 & pedersen\&07 \\
\hline
\end{tabular}

\medskip
$z$, L$_X$ and T$_X$ are redshift, core-excised X-ray temperature and core-excised soft band luminosity of the cluster from \citet{Mantz10}. kT ref. gives the reference for temperatures in \citet{Mantz10}, (1) are ASCA temperatures from \citet{Horner01}, 2 and 3 are Chandra temperatures from \citet{Vikhlinin09} and \citet{Mantz10}. $M_\mathrm{500}$, $M_\mathrm{200}$ and author code the spherical overdensity masses with respect to the critical density and author code fields in the LC$^2$ catalog of \citet{Sereno14b}. Author code applegate+14 points to \citet{Applegate14}, cypriano+04 to \citet{Cypriano04}, dahle06 to \citet{Dahle06}, kubo+09 to \citet{Kubo09}, mahdavi+13 to \citet{Mahdavi13}, okabe\&08 to \citet{Okabe08}, okabe+14b to \citet{Okabe14b}, pedersen\&07 to \citet{Pedersen07}, umetsu+09 to \citet{Umetsu09}, umetsu+11 to \citet{Umetsu11} and umetsu+14 to \citet{Umetsu14}. 
\end{table*}

\begin{table*}
\caption{\boldtext{The Eddington bias corrections for the literature high-mass sample.}}
\centering
\label{tab:litbias}
\begin{tabular}{lcccc}
\hline
Name & $ \frac{d \alpha(ln(M)}{d ln(M)}$ & M$_{rat}$ & T$_{X, rat}$ & L$_{X, rat}$ \\ 
\hline
Abell2029        & 1.879	 &     0.845 &       0.930	 &       0.595 \\
Abell478         & 2.244	 &     0.817 &       0.916	 &       0.538 \\
Abell2142        & 2.116	 &     0.827 &       0.921	 &       0.561 \\
Abell2244        & 1.234	 &     0.895 &       0.953	 &       0.710 \\
Abell2034        & 1.689	 &     0.859 &       0.935	 &       0.627 \\
Abell2204        & 3.163	 &     0.753 &       0.879	 &       0.417 \\
Abell2218        & 1.725	 &     0.856 &       0.933	 &       0.621 \\
Abell1914        & 1.818	 &     0.849 &       0.930	 &       0.604 \\
Abell665         & 2.282	 &     0.815 &       0.915	 &       0.533 \\
Abell520         & 1.854	 &     0.846 &       0.930	 &       0.603 \\
Abell963         & 1.635	 &     0.863 &       0.937	 &       0.640 \\
Abell1423        & 2.724	 &     0.783 &       0.888	 &       0.474 \\
Abell773         & 3.555	 &     0.727 &       0.867	 &       0.380 \\
Abell2261        & 3.193	 &     0.751 &       0.881	 &       0.419 \\
Abell1682        & 1.582	 &     0.867 &       0.868	 &       0.643 \\
Abell1763        & 2.665	 &     0.787 &       0.898	 &       0.483 \\
Abell2219        & 2.821     &     0.776 &       0.894	 &       0.462 \\
Abell2111        & 1.691	 &     0.859 &       0.927	 &       0.629 \\
Abell267         & 1.847	 &     0.847 &       0.923	 &       0.604 \\
Abell2390        & 2.757	 &     0.781 &       0.898	 &       0.472 \\
Abell1835        & 3.369	 &     0.739 &       0.878	 &       0.399 \\
Abell68          & 2.511	 &     0.798 &       0.889	 &       0.503 \\
Abell697         & 2.723	 &     0.783 &       0.888	 &       0.475 \\
Abell781         & 2.663	 &     0.787 &       0.880	 &       0.482 \\
Abell85          & 1.516	 &      0.872 &       0.943	 &       0.660 \\
Abell2597        & 1.185	 &     0.899 &       0.955	 &       0.720 \\
Abell1689        & 2.709	 &     0.784 &       0.899	 &       0.472 \\
Abell209         & 2.826	 &     0.776 &       0.889	 &       0.463 \\
Abell521         & 2.325	 &     0.811 &       0.912	 &       0.531 \\
Abell2537        & 2.325	&      0.811 &       0.901	 &       0.528 \\
MACSJ1115.8+0129 & 2.987	&      0.765 &       0.877	 &       0.443 \\
MACSJ0949.8+1708 & 2.751	&      0.781 &       0.847	 &       0.471 \\
MACSJ1731.6+2252 & 4.651	&      0.659 &       0.816	 &       0.280 \\
MACSJ2211.7-0349 & 3.500	&      0.730 &       0.815	 &       0.384 \\
MACSJ0429.6-0253 & 2.337	&      0.811 &       0.875	 &       0.527 \\
MACSJ1206.2-0847 & 3.236	&      0.748 &       0.863	 &       0.412 \\
MACSJ0417.5-1154 & 4.661	&      0.658 &       0.809	 &       0.279 \\
MACSJ2243.3-0935 & 4.569	&      0.664 &       0.816	 &       0.287 \\
RXJ0439.0+0715   & 2.546	&      0.795 &       0.901	 &       0.500 \\
Zwicky5247       & 1.176   &       0.899 &       0.932	 &       0.724 \\
Zwicky2089       & 1.439	&      0.879 &       0.911	 &       0.673 \\
Zwicky3146       & 3.142	&      0.754 &       0.883	 &       0.424 \\
\hline
\end{tabular}
\medskip

$ \frac{d \alpha(ln(M)}{d ln(M)}$ is the slope of the mass function, M$_{rat}$, T$_{X, rat}$ and L$_{X, rat}$ are the ratio of the Eddington bias corrected mass, temperature and luminosity to the uncorrected values.
\end{table*}

\label{lastpage}

\begin{thebibliography}{}

\bibitem[\protect\citeauthoryear{Allevato et 
al.}{2012}]{Allevato12} Allevato V., et al., 2012, ApJ, 758, 47 

\bibitem[\protect\citeauthoryear{Applegate et 
al.}{2014}]{Applegate14} Applegate D.~E., et al., 2014, MNRAS, 439, 

\bibitem[\protect\citeauthoryear{Bartelmann}{1996}]{1996A&A...313..697B} Bartelmann M., 1996, A\&A, 313, 697 

\bibitem[\protect\citeauthoryear{Becker 
\& Kravtsov}{2011}]{Becker11} Becker M.~R., Kravtsov A.~V., 2011, ApJ, 740, 25 

\bibitem[\protect\citeauthoryear{Ben{\'{\i}}tez}{2000}]{Benitez00} Ben{\'{\i}}tez N., 2000, ApJ, 536, 571 

\bibitem[\protect\citeauthoryear{Benjamin et 
al.}{2013}]{Benjamin13} Benjamin J., et al., 2013, MNRAS, 431, 
1547 

\bibitem[\protect\citeauthoryear{Bharadwaj et 
al.}{2015}]{Bharadwaj14} Bharadwaj V., Reiprich T.~H., Lovisari L., Eckmiller H.~J., 2015, A\&A, 573, AA75 

\bibitem[\protect\citeauthoryear{Bielby et 
al.}{2010}]{Bielby10} Bielby R.~M., et al., 2010, A\&A, 523, A66 

\bibitem[\protect\citeauthoryear{B{\"o}hringer et 
al.}{2007}]{Bohringer07} B{\"o}hringer H., et al., 2007, A\&A, 469, 363 

\bibitem[\protect\citeauthoryear{Connor et al.}{2014}]{Connor14} 
Connor T., Donahue M., Sun M., Hoekstra H., Mahdavi A., Conselice C.~J., 
McNamara B., 2014, ApJ, 794, 48 

\bibitem[\protect\citeauthoryear{Corless 
\& King}{2007}]{Corless07} Corless V.~L., King L.~J., 2007, MNRAS, 380, 149 

\bibitem[\protect\citeauthoryear{Cypriano et 
al.}{2004}]{Cypriano04} Cypriano E.~S., Sodr{\'e} L., Jr., Kneib 
J.-P., Campusano L.~E., 2004, ApJ, 613, 95 

\bibitem[\protect\citeauthoryear{Dahle}{2006}]{Dahle06} Dahle 
H., 2006, ApJ, 653, 954 

\bibitem[\protect\citeauthoryear{Donahue et al.}{2014}]{Donahue14} Donahue M., et al., 2014, arXiv, arXiv:1405.7876 

\bibitem[\protect\citeauthoryear{Duffy et al.}{2008}]{Duffy08} 
Duffy A.~R., Schaye J., Kay S.~T., Dalla Vecchia C., 2008, MNRAS, 390, L64 

\bibitem[\protect\citeauthoryear{Dutton 
\& Macci{\`o}}{2014}]{DM14} Dutton A.~A., Macci{\`o} A.~V., 2014, MNRAS, 441, 3359 

\bibitem[\protect\citeauthoryear{Eckmiller, Hudson, 
\& Reiprich}{2011}]{Eckmiller11} Eckmiller H.~J., Hudson D.~S., Reiprich T.~H., 2011, A\&A, 535, A105 

\bibitem[\protect\citeauthoryear{Eddington}{1913}]{Eddington13} 
Eddington A.~S., 1913, MNRAS, 73, 359 

\bibitem[\protect\citeauthoryear{Eisenstein et 
al.}{2001}]{2001AJ....122.2267E} Eisenstein D.~J., et al., 2001, AJ, 122, 
2267 

\bibitem[\protect\citeauthoryear{Erben et al.}{2013}]{Erben13} 
Erben T., et al., 2013, MNRAS, 433, 2545 

\bibitem[\protect\citeauthoryear{Fabjan et al.}{2011}]{Fabjan11} 
Fabjan D., Borgani S., Rasia E., Bonafede A., Dolag K., Murante G., 
Tornatore L., 2011, MNRAS, 416, 801 

\bibitem[\protect\citeauthoryear{Finoguenov, Reiprich 
\& B{\"o}hringer}{2001}]{Finoguenov01} Finoguenov A., Reiprich T.~H., B{\"o}hringer H., 2001, A\&A, 368, 749 

\bibitem[\protect\citeauthoryear{Finoguenov, B{\"o}hringer, 
\& Zhang}{2005}]{finoguenov05} Finoguenov A., B{\"o}hringer H., Zhang Y.-Y., 2005, A\&A, 442, 827 

\bibitem[\protect\citeauthoryear{Finoguenov et al.}{2007a}]{Finoguenov07} Finoguenov A., et al., 2007, ApJS, 172, 
182 

\bibitem[\protect\citeauthoryear{Finoguenov et 
al.}{2007b}]{Finoguenov07b} Finoguenov A., Ponman T.~J., Osmond 
J.~P.~F., Zimer M., 2007, MNRAS, 374, 737 

\bibitem[\protect\citeauthoryear{Finoguenov et 
al.}{2009}]{Finoguenov09} Finoguenov A., et al., 2009, ApJ, 704, 564 

\bibitem[\protect\citeauthoryear{Ford et al.}{2015}]{Ford15} 
Ford J., et al., 2015, MNRAS, 447, 1304 

\bibitem[\protect\citeauthoryear{Ford et al.}{2012}]{Ford12} 
Ford J., et al., 2012, ApJ, 754, 143 

\bibitem[\protect\citeauthoryear{George et al.}{2011}]{George11} 
George M.~R., et al., 2011, ApJ, 742, 125 

\bibitem[\protect\citeauthoryear{Giodini et 
al.}{2010}]{Giodini10} Giodini S., et al., 2010, ApJ, 714, 218 

\bibitem[\protect\citeauthoryear{Giodini et al.}{2013}]{Giodini13} Giodini S., Lovisari L., Pointecouteau E., 
Ettori S., Reiprich T.~H., Hoekstra H., 2013, SSRv, 177, 247 

\bibitem[\protect\citeauthoryear{Gozaliasl et 
al.}{2014}]{Gozaliasl14} Gozaliasl G., et al., 2014, A\&A, 566, AA140 

\bibitem[\protect\citeauthoryear{Heymans et 
al.}{2013}]{Heymans13} Heymans C., et al., 2013, MNRAS, 432, 2433 

\bibitem[\protect\citeauthoryear{Heymans et 
al.}{2012}]{Heymans12} Heymans C., et al., 2012, MNRAS, 427, 146 

\bibitem[\protect\citeauthoryear{Hamana, Takada, 
\& Yoshida}{2004}]{Hamana04} Hamana T., Takada M., Yoshida N., 2004, MNRAS, 350, 893 

\bibitem[\protect\citeauthoryear{Hildebrandt et 
al.}{2012}]{Hildebrandt12} Hildebrandt H., et al., 2012, MNRAS, 421, 
2355

\bibitem[\protect\citeauthoryear{Hirata et al.}{2004}]{Hirata04} 
Hirata C.~M., et al., 2004, MNRAS, 353, 529 

\bibitem[\protect\citeauthoryear{Hoekstra et 
al.}{2015}]{Hoekstra15} Hoekstra H., Herbonnet R., Muzzin A., 
Babul A., Mahdavi A., Viola M., Cacciato M., 2015, MNRAS in press, arXiv:1502.01883 


\bibitem[\protect\citeauthoryear{Hoekstra et 
al.}{2013}]{Hoekstra13} Hoekstra H., Bartelmann M., Dahle H., 
Israel H., Limousin M., Meneghetti M., 2013, SSRv, 177, 75 

\bibitem[\protect\citeauthoryear{Hoekstra et al.}{2012}]{Hoekstra12} Hoekstra H., Mahdavi A., Babul A., 
Bildfell C., 2012, MNRAS, 427, 1298 

\bibitem[\protect\citeauthoryear{Hoekstra et 
al.}{2011}]{Hoekstra11b} Hoekstra H., Hartlap J., Hilbert S., van 
Uitert E., 2011, MNRAS, 412, 2095 

\bibitem[\protect\citeauthoryear{Hoekstra, Yee, 
\& Gladders}{2004}]{Hoekstra04} Hoekstra H., Yee H.~K.~C., Gladders M.~D., 2004, ApJ, 606, 67 

\bibitem[\protect\citeauthoryear{Hoekstra et 
al.}{2001}]{Hoekstra01a} Hoekstra H., et al., 2001, ApJ, 548, L5 

\bibitem[\protect\citeauthoryear{Hoekstra}{2001}]{Hoekstra01b} Hoekstra H., 2001, A\&A, 370, 743 

\bibitem[\protect\citeauthoryear{Horner}{2001}]{Horner01} Horner 
D.~J., 2001, PhDT,  

\bibitem[\protect\citeauthoryear{Hudson et 
al.}{2010}]{Hudson10} Hudson D.~S., Mittal R., Reiprich T.~H., Nulsen P.~E.~J., Andernach H., Sarazin C.~L., 2010, A\&A, 513, A37 

\bibitem[\protect\citeauthoryear{Hudson et al.}{2013}]{Hudson13} 
Hudson M.~J., et al., 2013, arXiv, arXiv:1310.6784 

\bibitem[\protect\citeauthoryear{Israel et al.}{2014a}]{Israel14i} Israel H., Reiprich T.~H., Erben T., Massey R.~J., Sarazin C.~L., Schneider P., Vikhlinin A., 2014, A\&A, 564, A129 

\bibitem[\protect\citeauthoryear{Israel et al.}{2014b}]{Israel14ii} 
Israel H., Schellenberger G., Nevalainen J., Massey R., Reiprich T., 2014, 
arXiv, arXiv:1408.4758 


\bibitem[\protect\citeauthoryear{Jee et al.}{2011}]{Jee11} 
Jee M.~J., et al., 2011, ApJ, 737, 59 

\bibitem[\protect\citeauthoryear{Johnson et 
al.}{2011}]{Johnson11} Johnson R., Finoguenov A., Ponman T.~J., 
Rasmussen J., Sanderson A.~J.~R., 2011, MNRAS, 413, 2467 


\bibitem[\protect\citeauthoryear{Kaiser}{1986}]{Kaiser86} Kaiser 
N., 1986, MNRAS, 222, 323 

\bibitem[\protect\citeauthoryear{Kelly}{2007}]{Kelly07} Kelly 
B.~C., 2007, ApJ, 665, 1489 

\bibitem[\protect\citeauthoryear{Kettula, Nevalainen, 
\& Miller}{2013}]{Kettula13b} Kettula K., Nevalainen J., Miller E.~D., 2013, A\&A, 552, A47 

\bibitem[\protect\citeauthoryear{Kettula et 
al.}{2013}]{Kettula13} Kettula K., et al., 2013, ApJ, 778, 74 

\bibitem[\protect\citeauthoryear{Kilbinger et 
al.}{2013}]{Kilbinger13} Kilbinger M., et al., 2013, MNRAS, 430, 
2200 

\bibitem[\protect\citeauthoryear{Kitching et 
al.}{2014}]{Kitching14} Kitching T.~D., et al., 2014, arXiv, 
arXiv:1401.6842 

\bibitem[\protect\citeauthoryear{Kubo et al.}{2009}]{Kubo09} 
Kubo J.~M., et al., 2009, ApJ, 702, L110 

\bibitem[\protect\citeauthoryear{Le Brun et 
al.}{2014}]{LeBrun14} Le Brun A.~M.~C., McCarthy I.~G., Schaye 
J., Ponman T.~J., 2014, MNRAS, 441, 1270 

\bibitem[\protect\citeauthoryear{Leauthaud et 
al.}{2010}]{Leauthaud10} Leauthaud A., et al., 2010, ApJ, 709, 97 

\bibitem[\protect\citeauthoryear{Lovisari, Reiprich, 
\& Schellenberger}{2015}]{Lovisari15} Lovisari L., Reiprich T., Schellenberger G., 2015, A\&A, 573, AA118 

\bibitem[\protect\citeauthoryear{Mahdavi et 
al.}{2014}]{Mahdavi14} Mahdavi A., Hoekstra H., Babul A., 
Bildfell C., Jeltema T., Henry J.~P., 2014, ApJ, 794, 175 

\bibitem[\protect\citeauthoryear{Mahdavi et 
al.}{2013}]{Mahdavi13} Mahdavi A., Hoekstra H., Babul A., 
Bildfell C., Jeltema T., Henry J.~P., 2013, ApJ, 767, 116 

\bibitem[\protect\citeauthoryear{Mahdavi et 
al.}{2008}]{Mahdavi08} Mahdavi A., Hoekstra H., Babul A., Henry 
J.~P., 2008, MNRAS, 384, 1567 

\bibitem[\protect\citeauthoryear{Mandelbaum et 
al.}{2005}]{Mandelbaum05} Mandelbaum R., et al., 2005, MNRAS, 361, 
1287 

\bibitem[\protect\citeauthoryear{Mantz et al.}{2014}]{Mantz14} 
Mantz A.~B., et al., 2014, arXiv, arXiv:1407.4516 

\bibitem[\protect\citeauthoryear{Mantz et al.}{2010}]{Mantz10} 
Mantz A., Allen S.~W., Ebeling H., Rapetti D., Drlica-Wagner A., 2010, 
MNRAS, 406, 1773 

\bibitem[\protect\citeauthoryear{McCarthy et 
al.}{2010}]{Mccarthy10} McCarthy I.~G., et al., 2010, MNRAS, 406, 
822 

\bibitem[\protect\citeauthoryear{Meneghetti et 
al.}{2010}]{Meneghetti10} Meneghetti M., Rasia E., Merten J., Bellagamba F., Ettori S., Mazzotta P., Dolag K., Marri S., 2010, A\&A, 514, A93 

\bibitem[\protect\citeauthoryear{Miller et al.}{2013}]{Miller13} 
Miller L., et al., 2013, MNRAS, 429, 2858 

\bibitem[\protect\citeauthoryear{Miller et al.}{2007}]{Miller07} 
Miller L., Kitching T.~D., Heymans C., Heavens A.~F., van Waerbeke L., 
2007, MNRAS, 382, 315 

\bibitem[\protect\citeauthoryear{Miniati}{2015}]{Miniati15} 
Miniati F., 2015, ApJ, 800, 60 

\bibitem[\protect\citeauthoryear{Mirkazemi et 
al.}{2015}]{Mirkazemi15} Mirkazemi M., et al., 2015, ApJ, 799, 60 

\bibitem[\protect\citeauthoryear{Nagai, Kravtsov, 
\& Vikhlinin}{2007}]{Nagai07} Nagai D., Kravtsov A.~V., Vikhlinin A., 2007, ApJ, 668, 1 

\bibitem[\protect\citeauthoryear{Navarro, Frenk, 
\& White}{1997}]{NFW97} Navarro J.~F., Frenk C.~S., White S.~D.~M., 1997, ApJ, 490, 493 

\bibitem[\protect\citeauthoryear{Nevalainen, David, 
\& Guainazzi}{2010}]{Nevalainen10} Nevalainen J., David L., Guainazzi M., 2010, A\&A, 523, A22 

\bibitem[\protect\citeauthoryear{Okabe 
\& Umetsu}{2008}]{Okabe08} Okabe N., Umetsu K., 2008, PASJ, 60, 345 

\bibitem[\protect\citeauthoryear{Okabe et al.}{2010}]{Okabe10} 
Okabe N., Zhang Y.-Y., Finoguenov A., Takada M., Smith G.~P., Umetsu K., 
Futamase T., 2010, ApJ, 721, 875 

\bibitem[\protect\citeauthoryear{Okabe et al.}{2014}]{Okabe14b} 
Okabe N., et al., 2014, PASJ, 66, 99 

\bibitem[\protect\citeauthoryear{Parker et al.}{2005}]{Parker05} 
Parker L.~C., Hudson M.~J., Carlberg R.~G., Hoekstra H., 2005, ApJ, 634, 
806 

\bibitem[\protect\citeauthoryear{Pedersen 
\& Dahle}{2007}]{Pedersen07} Pedersen K., Dahle H., 2007, ApJ, 667, 26 

\bibitem[\protect\citeauthoryear{Pike et al.}{2014}]{Pike14} 
Pike S.~R., Kay S.~T., Newton R.~D.~A., Thomas P.~A., Jenkins A., 2014, 
arXiv, arXiv:1409.0723 

\bibitem[\protect\citeauthoryear{Planck Collaboration XXIX}{2013}]{PlanckXXIX} Planck Collaboration, et al., 2013, arXiv, 
arXiv:1303.5089 

\bibitem[\protect\citeauthoryear{Planelles et 
al.}{2014}]{Planelles14} Planelles S., Borgani S., Fabjan D., 
Killedar M., Murante G., Granato G.~L., Ragone-Figueroa C., Dolag K., 2014, 
MNRAS, 438, 195 

\bibitem[\protect\citeauthoryear{Poole et al.}{2007}]{Poole07} 
Poole G.~B., Babul A., McCarthy I.~G., Fardal M.~A., Bildfell C.~J., Quinn 
T., Mahdavi A., 2007, MNRAS, 380, 437 

\bibitem[\protect\citeauthoryear{Pratt et 
al.}{2007}]{pratt07} Pratt G.~W., B{\"o}hringer H., Croston J.~H., Arnaud M., Borgani S., Finoguenov A., Temple R.~F., 2007, A\&A, 461, 71 

\bibitem[\protect\citeauthoryear{Rasia et al.}{2012}]{Rasia12} 
Rasia E., et al., 2012, NJPh, 14, 055018 

\bibitem[\protect\citeauthoryear{Read, Guainazzi, 
\& Sembay}{2014}]{Read14} Read A.~M., Guainazzi M., Sembay S., 2014, A\&A, 564, A75 

\bibitem[\protect\citeauthoryear{Rykoff et al.}{2008}]{Rykoff08} 
Rykoff E.~S., et al., 2008, MNRAS, 387, L28 

\bibitem[\protect\citeauthoryear{Schaye et al.}{2010}]{Schaye10} 
Schaye J., et al., 2010, MNRAS, 402, 1536 

\bibitem[\protect\citeauthoryear{Schellenberger et 
al.}{2014}]{Schellenberger14} Schellenberger G., Reiprich T.~H., 
Lovisari L., Nevalainen J., David L., 2014, arXiv, arXiv:1404.7130 

\bibitem[\protect\citeauthoryear{Schmidt et 
al.}{2012}]{Schmidt12} Schmidt F., Leauthaud A., Massey R., 
Rhodes J., George M.~R., Koekemoer A.~M., Finoguenov A., Tanaka M., 2012, 
ApJ, 744, LL22 

\bibitem[\protect\citeauthoryear{Scoville et 
al.}{2007}]{Scoville07} Scoville N., et al., 2007, ApJS, 172, 1 

\bibitem[\protect\citeauthoryear{Semboloni et 
al.}{2011}]{Semboloni11} Semboloni E., Hoekstra H., Schaye J., van 
Daalen M.~P., McCarthy I.~G., 2011, MNRAS, 417, 2020 

\bibitem[\protect\citeauthoryear{Semboloni, Hoekstra, 
\& Schaye}{2013}]{Semboloni13} Semboloni E., Hoekstra H., Schaye J., 2013, MNRAS, 434, 148 

\bibitem[\protect\citeauthoryear{Sereno 
\& Ettori}{2014}]{Sereno14} Sereno M., Ettori S., 2014, arXiv, arXiv:1407.7868 

\bibitem[\protect\citeauthoryear{Sereno}{2014}]{Sereno14b} Sereno 
M., 2014, arXiv, arXiv:1409.5435 

\bibitem[\protect\citeauthoryear{Shaw et al.}{2010}]{Shaw10} 
Shaw L.~D., Nagai D., Bhattacharya S., Lau E.~T., 2010, ApJ, 725, 1452 

\bibitem[\protect\citeauthoryear{Shi 
\& Komatsu}{2014}]{Shi14} Shi X., Komatsu E., 2014, MNRAS, 442, 521 

\bibitem[\protect\citeauthoryear{Simpson et 
al.}{2013}]{Simpson13} Simpson F., et al., 2013, MNRAS, 429, 2249 

\bibitem[\protect\citeauthoryear{Smith et al.}{2005}]{Smith05} 
Smith G.~P., Kneib J.-P., Smail I., Mazzotta P., Ebeling H., Czoske O., 
2005, MNRAS, 359, 417 

\bibitem[\protect\citeauthoryear{Snowden et 
al.}{2008}]{Snowden08} Snowden S.~L., Mushotzky R.~F., Kuntz K.~D., Davis D.~S., 2008, A\&A, 478, 615 

\bibitem[\protect\citeauthoryear{Stanek et al.}{2010}]{Stanek10} 
Stanek R., Rasia E., Evrard A.~E., Pearce F., Gazzola L., 2010, ApJ, 715, 
1508 

\bibitem[\protect\citeauthoryear{Sun et al.}{2009}]{Sun09} 
Sun M., Voit G.~M., Donahue M., Jones C., Forman W., Vikhlinin A., 2009, 
ApJ, 693, 1142 

\bibitem[\protect\citeauthoryear{Umetsu et al.}{2014}]{Umetsu14} 
Umetsu K., et al., 2014, ApJ, 795, 163 

\bibitem[\protect\citeauthoryear{Umetsu et al.}{2011}]{Umetsu11} 
Umetsu K., Broadhurst T., Zitrin A., Medezinski E., Hsu L.-Y., 2011, ApJ, 
729, 127 

\bibitem[\protect\citeauthoryear{Umetsu et al.}{2009}]{Umetsu09} 
Umetsu K., et al., 2009, ApJ, 694, 1643 

\bibitem[\protect\citeauthoryear{van Daalen et 
al.}{2011}]{vanDaalen11} van Daalen M.~P., Schaye J., Booth C.~M., 
Dalla Vecchia C., 2011, MNRAS, 415, 3649 

\bibitem[\protect\citeauthoryear{van den Bosch}{2002}]{vandenBosch02} 
van den Bosch F.~C., 2002, MNRAS, 331, 98 

\bibitem[\protect\citeauthoryear{van Uitert et 
al.}{2011}]{vanUitert11} van Uitert E., Hoekstra H., Velander M., Gilbank D.~G., Gladders M.~D., Yee H.~K.~C., 2011, A\&A, 534, A14 

\bibitem[\protect\citeauthoryear{Velander et 
al.}{2014}]{Velander13} Velander M., et al., 2014, MNRAS, 437, 
2111 

\bibitem[\protect\citeauthoryear{Vikhlinin et 
al.}{2009}]{Vikhlinin09} Vikhlinin A., et al., 2009, ApJ, 692, 1033 

\bibitem[\protect\citeauthoryear{von der Linden et 
al.}{2014a}]{vdl14} von der Linden A., et al., 2014, MNRAS, 
443, 1973 

\bibitem[\protect\citeauthoryear{von der Linden et 
al.}{2014b}]{vdl14b} von der Linden A., et al., 2014, MNRAS, 
439, 2 

\bibitem[\protect\citeauthoryear{Wright 
\& Brainerd}{2000}]{Wright00} Wright C.~O., Brainerd T.~G., 2000, ApJ, 534, 34 

\bibitem[\protect\citeauthoryear{Zhang et 
al.}{2011}]{Zhang11} Zhang Y.-Y., Andernach H., Caretta C.~A., Reiprich T.~H., B{\"o}hringer H., Puchwein E., Sijacki D., Girardi M., 2011, A\&A, 526, A105 


\end{thebibliography}
\end{document}